\documentclass[10pt]{article}
\DeclareMathAlphabet{\scr}{U}{rsfs}{m}{n}
\usepackage{latexsym}
\usepackage[mathscr]{eucal}
\usepackage{amsfonts}
\usepackage{amscd}
\usepackage{amssymb}
\usepackage[centertags]{amsmath}
\usepackage{dsfont}
\usepackage{yfonts} 
\usepackage{enumerate}

\newcommand{\cleqn}{\setcounter{equation}{0}}
\newcommand{\newc}{\newcommand}
\newc{\eps}{\epsilon}
\newc{\lam}{\lambda}
\newc{\Lam}{\Lambda}
\newc{\ra}{\rightarrow}
\newc{\lra}{\leftrightarrow}
\newc{\wtilde}{\widetilde}
\newc{\ie}{{\it i.e.}}
\newc{\eg}{{\it e.g.}}
\newc{\rpv}{\not\!\! M_p}
\newc{\lsim}{\stackrel{<}{\sim}}
\newc{\beq}{\begin{equation}}
\newc{\eeq}{\end{equation}}
\newc{\beqn}{\begin{eqnarray}}
\newc{\eeqn}{\end{eqnarray}}
\newc{\PLB}{\emph{Phys.Lett.}{\bf{B}}}
\newc{\NPB}{\emph{Nucl.Phys.}{\bf{B}}}
\newc{\mcal}{\mathcal}
\newc{\bsym}{\boldsymbol}
\newc{\nonum}{\nonumber}
\newc{\ol}{\overline}
\begin{document}
\title{\hfill ~\\[-9mm]
       \hfill\mbox{\small UFIFT-HEP-07-14}\\[18mm]
\textbf{Dirac neutrinos and anomaly-free\\[2mm]discrete gauge symmetries\\[1mm]~}}
\author{Christoph Luhn$^{1}$,\footnote{
    E-mail: {\tt luhn@phys.ufl.edu}}
  ~~~ Marc Thormeier$^{2}$\footnote{
    E-mail: {\tt thor@th.physik.uni-bonn.de}}
  \\~\\~\\
 \emph{$^1${}~Institute for Fundamental Theory, Department of Physics,}\\
  \emph{University of Florida, Gainesville, FL 32611, USA}\\~\\
\emph{$^2${}~Physikalisches Institut der Universit\"at Bonn,}\\
  \emph{Nu\ss{}allee 12, 53115 Bonn, Germany}}\date{}
\maketitle

\begin{abstract}
\vspace{.2cm}\noindent
Relying on Dirac neutrinos allows an infinity of
anomaly-free discrete gauge symmetries to be imposed on the Supersymmetric
Standard Model, some of which are GUT-compatible.
\\

\end{abstract}

\section*{A few introductory words}
\cleqn

Dirac neutrinos have not (yet) been ruled  out by experiment, see \emph{e.g.}
Ref.~\cite{Strumia:2006db}. Provided a
satisfactory explanation for the smallness of their masses, see \emph{e.g.}
Refs.~\cite{Cleaver:1997nj,Langacker:1998ut,Gogoladze:2001kj,Hung:2002qp,Gherghetta:2003he,Abel:2004tt,Davoudiasl:2005ks,Gabriel:2006ns,Nandi:2007cw,Demir:2007dt}, 
they are an interesting alternative to the standard Majorana picture, in
particular concerning cosmological aspects, see \emph{e.g.}
Refs.~\cite{Dick:1999je,Asaka:2005cn,Gu:2006dc,Gu:2007mi,Gu:2007mc,Gu:2007gy}.
For Dirac neutrinos, renormalization group effects have been studied in
Ref.~\cite{Lindner:2005as}, and the possibility of having zero textures in the
mass matrix  of Dirac neutrinos  has been considered in
Ref.~\cite{Hagedorn:2005kz}.  

It is the purpose  of this letter
to draw the attention to yet another of their features: We will show that the
assumption of Dirac rather than Majorana neutrino masses is consistent with
infinitely many mutually non-equivalent discrete gauge symmetries (DGSs). They
are so-called anomaly-free, and can be imposed on the Supersymmetric Standard
Model (SSM) to forbid unwanted, \emph{e.g.} proton endangering, operators. 
The possibilities thus far, namely
baryon triality~($B_3$), recently introduced proton hexality~($P_6$) and
well-known matter parity~($M_p$),  {\it cf.}
Refs.~\cite{Ibanez:1991hv,Ibanez:1991pr,Dreiner:2005rd}, 
are based on Majorana neutrino masses.

The content is as follows. In Sect.~\ref{discretesect} and
Appendix~\ref{endlichregen!} we review Abelian discrete symmetries; in
Sect.~\ref{section3} and
Apps.~\mbox{\ref{stringkacke},\ref{ggT},\ref{singclas},\ref{casestudy}}
we describe how DGSs emerge from a high energy theory. Sect.~\ref{mainpart}
constitutes the main part of this text, determining  the anomaly-free DGSs
which rely on Dirac neutrinos. Sect.~\ref{physicssetc} discusses the physical
implications of the various newly-found DGSs; Sect.~\ref{GUTcomp} together
with Appendix~\ref{gutapp} analyzes whether the corresponding discrete charges are
compatible with unification. In Sect.~\ref{toym}, as an example, we present 
toy models of how
 $\mathds{Z}_4$-DGSs might arise. Sect.~\ref{dsco} concludes.

%%%%%%%%%%%%%%%%%%%%%%%%

%%%%%%%%%%%%%%%%%%%%%%%%

%%%%%%%%%%%%%%%%%%%%%%%%

\section{\label{discretesect}Discrete symmetries}
\cleqn

Thus far no processes  which are lepton and/or baryon number violating have
been observed, which is why the corresponding renormalizable and
non-renormalizable operators in the Lagrangian density~$\scr{L}$ have to be
either strongly suppressed
or absent altogether.\footnote{For instance, the
  renormalizable superpotential operators $QL\overline{D}$ and
  $\overline{UDD}$ together cause rapid proton decay if neither of them is
  very strongly suppressed. But also \emph{e.g.} $QQQL$, though
  non-renormalizable and thus suppressed by presumably the gravitational
  scale, may cause havoc to the proton. So one needs help from \emph{e.g.} 
family symmetries,  see for instance Refs.~\cite{Dreiner:2003hw,Dreiner:2003yr,Harnik:2004yp,Larson:2004ji}.}    
The latter can be readily obtained by relying on a discrete symmetry (DS): One
demands~$\scr{L}$ to be invariant under a discrete  transformation of the
fields, $\varphi~\rightarrow~ \hat{O}^\mathrm{DS}_\varphi\cdot\varphi$. In
case the DS is a $\mathds{Z}_N$-symmetry, with $N\in\{2,3,...\}$, this reads
$\hat{O}^\mathrm{DS}_\varphi   =\mathrm{e}^{\frac{\mathrm{2\pi i}}{N}\cdot
  z_\varphi}$, with $z_\varphi\in\{0,...,N-1\}$. $\mathds{Z}_2$-symmetries
are commonly labeled ``parities'', Grossman and Haber \cite{Grossman:1998py}
coined the word ``triality'' for a  $\mathds{Z}_3$-symmetry, and in
Ref.~\cite{Dreiner:2005rd} $\mathds{Z}_6$-symmetries  were called
``hexalities''.  We will encounter further ``$N$-alities'' later on. Of course,
one can also have DSs which are not $\mathds{Z}_N$-symmetries (or direct
products thereof), which means that they are non-Abelian. In the following, we
will not concern ourselves with such non-Abelian DSs. 

Assuming the existence of all Standard Model (SM) gauge invariant operators in
the renormalizable and matter parity conserving superpotential (for notation see below\footnote{$Q$, $\overline{D}$, $\overline{U}$,  $L$,
  $\overline{E}$, $H_d$ and $H_u$ are the left-chiral superfields of the
  left-handed quark doublets, $d$-type and $u$-type antiquark singlets,
  left-handed lepton doublets, antilepton singlets and the two  Higgs
  doublets, respectively; the $h^{...}$ denote Yukawa matrices.  $SU(3)_C$,
  $SU(2)_W$ and generational indices are suppressed.}) 
\beq\label{mssmsuperpot}
\scr{W}_{\mathrm{MSSM}}=h^dQH_d\overline{D}+h^uQH_u\overline{U} +
h^eLH_d\overline{E}+\mu H_dH_u\ ,
\eeq
generation-independent  $\mathds{Z}_N$-symmetries can be classified
\cite{Ibanez:1991pr} according to which renormalizable lepton number
($n_{\mathrm{lepton}}$) and/or  quark number ($n_{\mathrm{quark}}$) violating
operators are forbidden: Allowing for the $n_{\mathrm{lepton}}$-violating
operator $QL\overline{D}$ automatically allows the existence of
$LL\overline{E}$ and $LH_u$ from the DS point of view, see Eq.~(\ref{A0}). Similarly, forbidding
one of these three terms automatically forbids all; such a statement
applies to  non-renormalizable operators as well. So the 
\mbox{(non-)}existence of one operator is accompanied by the
\mbox{(non-)}existence of a whole set of operators. Hence, concerning all lepton
and/or baryon number violating operators up to dimension five
\cite{Ibanez:1991pr,Dreiner:2005rd,Sakai:1981pk,Weinberg:1981wj,Allanach:2003eb} we find schematically, see Appendix~\ref{endlichregen!},
\begin{eqnarray} 
QL\overline{D}&\Longleftrightarrow &\Big\{QL\overline{D},\,LL\overline{E},\,
LH_u,\, Q\overline{UE}H_d, \nonumber \\
&&  ~~\ LH_uH_dH_u,\,  Q\overline{U}L^\dagger,\,
\overline{E}H_d{H_u}^\dagger,\, \overline{U}\overline{E}\overline{D}^\dagger
\Big\} \, ,\nonumber\\
\overline{UDD}& \Longleftrightarrow & \Big\{\overline{UDD},\, QQQH_d,\,
QQ\overline{D}^\dagger \Big\} \, ,\label{BVLVoperators}\\ 
QQQL&\Longleftrightarrow &\Big\{QQQL,\,\overline{UUDE} \Big\} \, ,\nonumber\\
LH_uLH_u & \Longleftrightarrow &  \Big\{LH_uLH_u   \Big\}\, .\nonumber
\end{eqnarray}
The terms on the left-hand side should be viewed as representatives for the
complete set on the right-hand side. [Lines~1, 3 and 4 will be extended in
Eq.~(\ref{BVLVwithN}).] The representatives of the  first two lines were used
by Ib\'a\~nez and Ross, see Ref.~\cite{Ibanez:1991pr}, to  classify the DSs in
terms of the operators $QL\overline{D}$ and $\overline{UDD}$, see also
Table~\ref{asasassa}: 
\begin{table}[t]
\begin{center}
\hspace{0mm}\begin{tabular}{||c||c|c||}
\hline\hline
$\phantom{\Bigg|}$ & $\not\!\exists~\overline{U}\overline{D}\overline{D}$ &
$\exists~\overline{U}\overline{D}\overline{D}$ \\ \hline\hline
$\phantom{\Bigg|}$ $\not\!\exists~ QL\overline{D}~~$ &   
\emph{matter} $N$-alities, \emph{e.g.} $M_p$, $P_6$ & 
\emph{lepton} $N$-alities, \emph{e.g.} $L_p$ \\ \hline
$\phantom{\Bigg|}$ $\exists~ QL\overline{D}~~$ &  
\emph{baryon} $N$-alities, \emph{e.g.} $B_p$, $B_3$ & no DS at all
\\\hline\hline  
\end{tabular}
\end{center}
\caption{\label{asasassa}The classification of different
  $\mathds{Z}_N$-symmetries \'a la Ib\'a\~nez and Ross, {\it cf.}
  Ref.~\cite{Ibanez:1991pr}.  
  For notations like $M_p$, $P_6$, \emph{etc.} see Table~\ref{commonDS}.} 
\end{table}
\begin{itemize}
\item A symmetry forbidding both operators is called a {\it matter $N$-ality} or \emph{generalized matter parity}. 
\item If only $\overline{U}\overline{D}\overline{D}$ is forbidden we speak of
  a {\it baryon $N$-ality} (\emph{generalized baryon parity}).
\item  Forbidding only $QL\overline{D}$ yields a {\it lepton $N$-ality} (\emph{generalized lepton parity}).
\item Allowing  both operators is compatible only with a
  $\mathds{Z}_N$-symmetry where the discrete charges are proportional to the
  hypercharge:\footnote{Solving $z_Q+z_{H_d}+z_{\overline{D}}=0$,
  $z_Q+z_{H_u}+z_{\overline{U}}=0$, $z_L+z_{H_d}+z_{\overline{E}}=0$,
  $z_{H_d}+z_{H_u}=0$, $z_Q+z_{L}+z_{\overline{D}}=0$ and
  $z_{\overline{U}}+2z_{\overline{D}}=0$ gives that $z_{...}\propto
  Y_{...}~\mathrm{mod}~N$.} $z_{...} \propto Y_{...}~\mathrm{mod}~N$. Such a
  DS is trivial.
\end{itemize}
One can already see that no {lepton} or {baryon} $N$-ality is
compatible with a Georgi-Glashow $SU(5)$, in which $QL\overline{D}$ and
$\overline{U}\overline{D}\overline{D}$  both originate from a
$\bf{10}~\overline{\bf{5}}~\overline{\bf{5}}$. Likewise proton hexality $P_6$
\cite{Dreiner:2005rd} is incompatible with $SU(5)$ as it allows
$Q\overline{U}Q\overline{D}$ but forbids $QQQL$, whereas both these operators
come from a $\bf{10}~\bf{10}~\bf{10}~\overline{\bf{5}}$.

The best-known example for a $\mathds{Z}_2$-symmetry is $R$-parity ($R_p$)
\cite{Farrar:1978xj}. 
The discrete charge $z^{R_p}_{\varphi}$ for a field $\varphi$  is given by  
$z^{R_p}_{\varphi}={n_{\mathrm{quark}}(\varphi)+n_{\mathrm{lepton}}(\varphi) +
  2\cdot s(\varphi)}$, $s$ being the spin. $R_p$ is defined for \emph{fields}
rather than \emph{superfields}, providing  a useful tool to classify whether a
particle is part of the \mbox{(2Higgs-)SM} or whether it is a superpartner of 
one of these, \emph{i.e.} whether it has a supersymmetric motivation. If one
demands invariance of the Lagrangian density $\scr{L}$ under $R_p$
($\scr{L}_{R_p}$), all lepton and baryon number violating
\emph{renormalizable} operators are forbidden.  

$R_p$ can be modified to  $R_p^{\mathrm{susy}}$
\cite{Dimopoulos:1981dw,Bento:1987mu} which acts on whole 
superfields $\varPhi$ (rather than fields): $z^{R_p^{\mathrm{susy}}}_\varPhi =
{n_{\mathrm{quark}}(\varPhi)+n_{\mathrm{lepton}}(\varPhi)}$, constraining the
super- and K\"ahler potential such that the result is~$\scr{L}_{R_p}$. Other
examples of $\mathds{Z}_2$-symmetries are baryon parity, $B_p$,
$z^{B_p}_\varPhi  ={n_{\mathrm{quark}}(\varPhi)}$, and lepton parity, $L_p$,
$z^{L_p}_\varPhi  ={n_{\mathrm{lepton}}(\varPhi)}$. Table~\ref{commonDS}
summarizes these common DSs together with the ones found in
\cite{Ibanez:1991pr} and \cite{Dreiner:2005rd}.  
\begin{table}[t]
\begin{center}
\hspace{0mm}\begin{tabular}{||l||c||c|c|c|c|c|c|c||c||}
\hline\hline
$\phantom{\Bigg|}$ &$N$ & $Q$ & $\overline{D}$  & $\overline{U}$  & $L$ & $\overline{E}$ & $H_d$ & $H_u$ & comments\\
\hline
\hline
$Y/Y_Q$ & - & 1 & 2 & $-4\phantom{-}$ & $-3\phantom{-}$ & 6 & $-3\phantom{-}$ & $3$ & \\
\hline
\hline
$n_{\mathrm{quark}}$ &- &  1 & $-1\phantom{-}$ & $-1\phantom{-}$ & 0 & 0 & 0 & 0 & \\
\hline
$n_{\mathrm{lepton}}$ & - &  0 & $0$ & $0$ & $1$ & $-1\phantom{-}$ & 0 & 0 & \\
\hline
\hline
$B_p$  & 2& 1 & 1 &  1 & 0 &0 &0 &0 & \begin{tabular}{c}  
baryon
$N$-ality,\\anomalous\end{tabular} \\
\hline
${B_p}^\prime$  &  2&0 &1 &1 &1 &0 &1 &1 & \begin{tabular}{c}
baryon
$N$-ality, \\anomalous\end{tabular}\\
\hline
$L_p$  &  2&0 &0 &0 &1 &1 &0 &0 &  \begin{tabular}{c}
lepton
$N$-ality,\\anomalous\end{tabular}  \\
\hline
${L_p}^\prime$  & 2&1 &0 &0 &0 &1 &1 &1 & \begin{tabular}{c}
lepton
$N$-ality, \\anomalous\end{tabular}\\
\hline
$M_p\equiv {R_p^{\mathrm{susy}}}^\prime$  &  2&0& 1& 1& 0& 1& 1& 1&
\begin{tabular}{c} 
matter
$N$-ality,\\
Pati-Salam compatible
\end{tabular}   \\
\hline
${M_p}^\prime\equiv R_p^{\mathrm{susy}}$  &  2&1 &1 &1 &1 &1 &0 &0  &
\begin{tabular}{c}
matter
$N$-ality,\\
$SO(10)$ compatible \end{tabular}  \\
\hline
\hline
$B_3$  &  3&0 &1 &2 &2 &2 &2 &1  & 
baryon
$N$-ality\\
\hline
${B_3}^\prime$ &  3&1 &0 &1 &2 &2 &2 &1   & 
baryon
$N$-ality \\
\hline
${B_3}^{\prime\prime}$  &  3& 2 &2 &0 &2 &2 &2 &1  & 
baryon
$N$-ality \\
\hline\hline
$P_6$                                     & 6&0 &5 &1 &4 &1 &1 &5 &
\begin{tabular}{c} 
matter
$N$-ality,
\\ same as ${M_p}\times {B_3}$ \end{tabular}  \\
\hline
${P_6}^{\prime}$                          &6&1 &1 &3 &1 &1 &4 &2
&\begin{tabular}{c} 
matter
$N$-ality,
\\ same as ${M_p}^\prime\times {B_3}^{\prime\prime}$ \end{tabular}  \\
\hline
${P_6}^{\prime\prime}$                    &6&2 &3 &5 &4 &1 &1 &5
&\begin{tabular}{c} 
matter
$N$-ality,
\\ same as ${M_p}\times {B_3}^{\prime}$ \end{tabular}  \\
\hline
${P_6}^{\prime\prime\prime}$              &6&3 &5 &1 &1 &1 &4 &2
&\begin{tabular}{c} 
matter
$N$-ality,
\\ same as ${M_p}^\prime\times {B_3}^{}$ \end{tabular}  \\
\hline
${P_6}^{\prime\prime\prime\prime}$        &6&4 &1 &3 &4 &1 &1 &5
&\begin{tabular}{c} 
matter
$N$-ality,
\\ same as ${M_p}\times {B_3}^{\prime\prime}$ \end{tabular}  \\
\hline
${P_6}^{\prime\prime\prime\prime\prime}$  &6&5 &3 &5 &1 &1 &4 &2 &
\begin{tabular}{c} 
matter
$N$-ality, 
\\ same as ${M_p}^\prime\times {B_3}^{\prime}$ \end{tabular} \\
\hline\hline
\end{tabular}
\end{center}
\caption{\label{commonDS}Common DSs. The first line gives the hypercharges of
  the superfields $Q$, $\overline{D}$ {\it etc.}; the second and the third
  lines list the corresponding quark and lepton number; the other lines show
  the discrete charges of the superfields under various DSs.}
\end{table}
The primed DSs are obtained from the unprimed by so-called
``hypercharge-shifts'', see  Items~\mbox{\ref{itten},\ref{rmnp}} in
Sect.~\ref{section3} as well as  Ref.~\cite{Dreiner:2005rd}. The constraining
effects of each of $\{B_p,{B_p}^\prime\}$ are identical, likewise for
$\{L_p,{L_p}^\prime\}$, $\{M_p,{R_p^{\mathrm{susy}}}\}$ [even though \emph{e.g.}
${R_p^{\mathrm{susy}}}$ is compatible with $SO(10)$ whereas  $M_p$ is only
Pati-Salam compatible, see Section~\ref{GUTcomp}],
$\{B_3,{B_3}^\prime,{B_3}^{\prime\prime}\}$ and
$\{P_6,{P_6}^\prime,...,{P_6}^{\prime\prime\prime\prime\prime}\}$. Examining
the consequences of the $P_6$-symmetries, we find that they are very
restrictive; as was proposed in Ref.~\cite{Dreiner:2005rd}: "$P_6$ is
\emph{the} DS of the MSSM" if one relies on Majorana neutrinos.

%%%%%%%%%%%%%%%%%%%%%%%%%%%%%%%%

%%%%%%%%%%%%%%%%%%%%%%%%%%%%%%%%

%%%%%%%%%%%%%%%%%%%%%%%%%%%%%%%%

\section{\label{section3}Discrete \emph{gauge} symmetries}
\cleqn

It can be argued that global DSs are violated by quantum gravitational effects
\cite{Krauss:1988zc}, which at first sight renders the  use of DSs
impractical. There is however a loop-hole:   If the DS  is   a so-called
D\emph{gauge}S (DGS), \emph{i.e} if it is the  remnant/residual/left-over of a
spontaneously broken local  gauge symmetry, then no wormholes \emph{etc.}
screw up its  performance \cite{Banks:1989ag,Preskill:1990bm}. 
The underlying ``mother symmetry'' of course must
not cause trouble with anomalies, from which follows that not every DS is
automatically feasible; for instance, as we will see in Sect.~\ref{mainpart},
$B_p$, ${B_p}^\prime$, $L_p$ and ${L_p}^\prime$ of Table~\ref{commonDS} cannot
originate from an anomaly-free high-energy $U(1)$ symmetry. 

In what follows we shall in a top-down fashion describe how DSs arise from  a
local gauge symmetry at high energies, listing step-by-step which
transformations are performed and/or which assumptions are made, to finally
arrive at the discrete anomaly equations which are the starting point of
Sect.~\ref{mainpart}. We try to stay as general as possible as long as
possible.

Though the  local gauge symmetry could in principle be Abelian or non-Abelian,
$R$- or \mbox{non-$R$}, for the rest of this paper, we shall consider a single
non-$R$ local $U(1)$ gauge group. Hence we restrict our DGSs to be
$\mathds{Z}_N$-symmetries. 

\emph{Hasty readers who are familiar with this subject may want to jump 
ahead directly to the next section, assuming a non-anomalous ``mother'' 
$U(1)_X$, no SM-singlets except the  $U(1)_X$-breaking superfield
$\mathfrak{A}$ with $X$-charge $N$ and three generations of right-handed
neutrinos, all $X$-charges being integer numbers, the discrete charges 
of the MSSM superfields and the neutrinos being generation-independent, all
SM-charged matter which is beyond the MSSM being heavy.}  

We start with an $SU(3)_C\times SU(2)_W\times U(1)_{Y^\prime}\times
U(1)_{X^\prime}$-invariant quantum field theory supposedly coming from a
string;  $U(1)_{Y^\prime}$ is not yet to be identified with the
SM-hypercharge, because for the sake of generality we take both $U(1)$-factors
to be possibly anomalous at first. Of course, if instead one starts  
\emph{e.g.} with  an  $SU(5)\times U(1)_{X^\prime}$-invariant theory, some of
the  following points are obviously rendered moot, and other blatant steps 
[like the breaking of $SU(5)$] have to be introduced at obvious places. 
Up to short summaries of the corresponding points, we have relegated 
the first seven steps in which the K\"ahler potential is canonicalized, the
dilaton acquires a VEV and the anomaly is rotated into the $U(1)_X$ alone 
to Appendix~\ref{stringkacke}. %Hence our newly defined 
%$U(1)_Y$ is non-anomalous, and we can proceed with the following list.

\begin{enumerate}%\setcounter{enumi}{7}
\item The Ka\v{c}-Moody matrix of the two $U(1)$ factors is taken
to be positive-definite.
\item The Ka\v{c}-Moody matrix is diagonalized.
\item The effects of the $U(1)_X \times U(1)_Y$ transformations are discussed:
  anomalies and the dilaton-originated Green-Schwarz shift.
\item The two above-mentioned effects mutually cancel.
\item The dilaton acquires a VEV, generating Fayet-Iliopoulos as well as
kinetic gauge terms.
\item The kinetic gauge terms are canonicalized.
\item We rotate  such  that  all Fayet-Iliopoulos terms are condensed in just
one of the $U(1)$ factors.
\item \label{AundO}One demands that some left-chiral superfields $A_i$
(not to be confused with the anomaly coefficients $\mathcal A_{abc} =
\mathrm{Trace}[\{ T^a,\,T^b  \} \cdot T^c]$, the $T$'s being the gauge group
generators) and $\Omega_j$ are SM-uncharged but $X$-charged. The scalar
components of the $A_i$ shall later acquire vacuum expectation values (VEVs)
and thus play the role of Higgs fields for the $U(1)_X$; if $U(1)_X$ is
generation-dependent, the $A_i$ are sometimes called flavons. The $\Omega_j$
on the other hand  denote all other SM-singlets like \emph{e.g.} a
right-handed neutrino $\overline{\mathcal{N}}$.  

\item \label{yshift}Next, one requires for the $Y$-charges that
$Y_Q+Y_{H_d}+Y_{\overline{D}}=Y_Q+Y_{H_u}+Y_{\overline{U}}=Y_L+Y_{H_d}+Y_{\overline{E}}=0$. This,
together with the vanishing of the anomaly coefficients $\mathcal A_{CCY}$,
$\mathcal A_{WWY}$, $\mathcal A_{GGY}$ and the assumption that all SM-charged
matter beyond the MSSM is vectorlike, allows one to identify
$U(1)_Y$ with the hypercharge, its values given in Table~\ref{commonDS}. Note
that if a set of $X$-charges $X_i$  gives a certain value for the overall
$X$-charge of a $Y$-invariant operator, then the set $X_i+\alpha Y_i$, with
$\alpha\in\mathds{R}$, constitutes the same value. The replacement   
$X_i\rightarrow X_i+\alpha Y_i $ is the so-called $Y$- or hypercharge-shift,
parameterized by $\alpha$.\footnote{It is important to note that $Y$-shifted
  $X$-charges represent different high energy physics. Examples are  \emph{a)}
  cross sections depend explicitly on charges, \emph{b)} one set of
  $X$-charges may be $SO(10)$ compatible, unlike its $Y$-shifted set, see
  Item~\ref{rmnp} as well as Section~\ref{GUTcomp} with the parameter $r$,
  and \emph{c)} the beta-function of $g_X$ is Trace[$X^2$]-dependent and is
  thus sensitive to $Y$-shifts, see \emph{e.g.} Ref.~\cite{dasewigepaper}.}
Evidently, terms which are $U(1)_Y$-allowed and $U(1)_X$-forbidden/allowed are
also forbidden/allowed after a $Y$-shift of the $X$-charges. Also the four
linear anomalies $\mathcal{A}_{CCX}$, $\mathcal{A}_{WWX}$, $\mathcal{A}_{YYX}$,
$\mathcal{A}_{GGX}$ [{\it cf.}~Eq.~(\ref{x-eq})] remain invariant under this
shift, whereas\\[0mm]
\beq
\mathcal A_{XXX}\rightarrow \mathcal A_{XXX}+6\pi^2 \alpha^2 k_Y X_S\ ,\qquad
\mathcal A_{YXX}\rightarrow \mathcal A_{YXX}+4\pi^2 \alpha k_Y X_S\label{@@@}\
.~~~\eeq\\[-2mm]
Here, $k_Y$ is the Ka\v{c}-Moody level of $U(1)_Y$, and $X_S$ is a real
parameter introduced in Eq.~(\ref{s-trafo}). Only if 
$X_S=0$, all anomaly coefficients involving $U(1)_X$ are invariant
under $Y$-shifts. In this case, due to the Green-Schwarz anomaly cancellation
condition of Eq.~(\ref{x-eq}), $U(1)_X$ is non-anomalous. Therefore,
starting with an anomaly-free $U(1)_X$, the equations which constrain the
remnant $\mathds{Z}_N$-symmetry in Sect.~\ref{mainpart} are not changed by
$Y$-shifts. For an example of models related by a $Y$-shift see
Sect.~\ref{toym}.

\item \label{itten} 
One postulates: With two sets of integers $n_i$, ${n^\prime}_j$ fulfilling
$\sum_i n_i Y_i=\sum_j {n^\prime}_j Y_j=0$, all $\frac{\sum_i n_i X_i}{\sum_j
  {n^\prime}_j X_j}$, \emph{i.e.} all ratios of $X$-charges of terms which are
$Y$-invariant, are rational numbers. Moreover, instead of making this
operator-wise requirement, we demand in a field-wise fashion that the
$X$-charges are such that all $X_i/X_j$ are rational numbers (charge
quantization). If $X_S=0$, this more restrictive requirement could be weakened
to demanding that there is a $Y$-shift relating the original set of
$X$-charges to another set for which all $X_i/X_j$ are rational numbers; for
simplicity we shall \emph{not} stick to this option.

\item \label{rescale} The previous Item allows to rescale the $X$-charges such
that they all take their smallest possible integer values.

\item With $\varPhi_k$ now denoting any superfield which is not an $A_i$ (see
  Item~\ref{AundO}), any super- or K\"ahler potential term $T$ composed of
  $k_{\mathrm{max}}$ different species of superfields which is $SU(3)_C\times
  SU(2)_W \times U(1)_Y$ invariant can be written in the form 
  $T={\varPhi_1}^{n_{\varPhi_1}}\cdot{\varPhi_2}^{n_{\varPhi_2}}\cdot...\cdot{\varPhi_{k_{\mathrm{max}}}}^{n_{\varPhi_{k_{\mathrm{max}}}}}$,
  the $n_{...}$ being integer numbers, denoting how often the corresponding 
  superfield appears in the term; note that ``${\varPhi}^{-1}$'' means 
  ``${\varPhi}^{\dagger}$''. However, it is by far not guaranteed that
  $n_{\varPhi_1}\!\cdot X_{\varPhi_1}+\,n_{\varPhi_2}\!\cdot X_{\varPhi_2}+...+\,n_{\varPhi_{k_{\mathrm{max}}}}\!\cdot X_{\varPhi_{k_{\mathrm{max}}}}\!=\,0$. 
But suppose that the excess $X$-charge can be compensated by several
powers of the superfield $A_1$. In this case  
$
\widetilde{T}={A_1}^{-({n_{\varPhi_1}\!\cdot X_{\varPhi_1}+\,n_{\varPhi_2}\!\cdot X_{\varPhi_2}+...+\,n_{\varPhi_{k_{\mathrm{max}}}}\!\cdot X_{\varPhi_{k_{\mathrm{max}}}}})/{X_{A_1}}}\cdot T
$
is $U(1)_X$-invariant. If there are several $A_j$ with different $X$-charges,
it is for the purposes in this paper useful to work with an
``effective~$A$'' or ``reduced~$A$`` which we will label $\mathfrak{A}$. Taking into account the Giudice-Masiero/Kim-Nilles mechanism
\cite{Giudice:1988yz,Kim:1994eu}, its
$X$-charge is the greatest common divisor of the $X$-charges of all the~$A_j$,
see Appendix~\ref{ggT}. $\widetilde{T}$ then  generalizes to \\[1mm]
\beq\label{aHoch}
{\mathfrak{A}}^{-({n_{\varPhi_1}\!\cdot X_{\varPhi_1}+\,n_{\varPhi_2}\cdot
  X_{\varPhi_2}+...+\,n_{\varPhi_{k_{\mathrm{max}}}}\!\cdot
  X_{\varPhi_{k_{\mathrm{max}}}}})/{X_{\mathfrak{A}}}}~~\cdot~~{\varPhi_1}^{n_{\varPhi_1}}\cdot{\varPhi_2}^{n_{\varPhi_2}}\cdot...\cdot{\varPhi_{k_{\mathrm{max}}}}^{n_{\varPhi_{k_{\mathrm{max}}}}}\ .
\eeq\\[-2mm]
As an example, consider $X_{A_1}=\sqrt{13}$, $X_{A_2}=\frac{7}{2}\sqrt{13}$,
$X_{A_3}=\frac{7}{3}\sqrt{13}$ and $X_{Q^1}=\frac{1}{5} \sqrt{13}$,
  $X_{H_d}=\frac{1}{15}\sqrt{13}$, $X_{\overline{D^1}}=\frac{2}{5}\sqrt{13}$
as a starting point. Then rescale the $X$-charges such that all fields have
integer charges, thus multiply by $2\cdot3\cdot5/\sqrt{13}$, arriving at
$X_{A_1}=2\cdot3\cdot5=30$, $X_{A_2}=3\cdot5\cdot7=105$, 
$X_{A_3}=2\cdot5\cdot7=70$ and $X_{Q^1}=6$, $X_{H_d}=2$,
$X_{\overline{D^1}}=12$. The greatest common divisor of the $A_{1,2,3}$ is
thus $5$, so $|X_\mathfrak{A}|=5$ (so \emph{e.g.}
$\mathfrak{A}=A_2{A_1}^\dagger{A_3}^\dagger$). Therefore, as we will argue
  later on, one arrives at a $\mathds{Z}_5$-symmetry with $z_{Q^1}=1$, $z_{\overline{D^1}}=z_{H_d}=2$. For another example see Appendix~\ref{casestudy}.

\item There is however an important \emph{caveat} to Eq.~(\ref{aHoch}). As the
Hamiltonian density necessarily is a polynomial of fields
\cite{Weinberg:1995mt,Weinberg:1996kw}, in order to satisfy the cluster
decomposition principle (CDP) \cite{Wichmann:1963}, {\it i.e.} distant
experiments have uncorrelated results, one may only have integer exponents of
the fields. This then translates to the requirement that every super- and
K\"ahler potential term may contain only integer powers of the superfields,
dictating that $n_{\varPhi_1}\!\cdot X_{\varPhi_1}+\,n_{\varPhi_2}\!\cdot
X_{\varPhi_2}+...+\,n_{\varPhi_{k_{\mathrm{max}}}}\!\cdot
X_{\varPhi_{k_{\mathrm{max}}}}$ is an integer multiple of 
$X_{\mathfrak{A}}$, otherwise the whole term is forbidden.    
 
\item \label{VI} The $A_i$ and thus also $\mathfrak{A}$ acquire VEVs, so $U(1)_X$ is broken. It must be ensured at all costs that those terms which are
(phenomenologically) desired  have $X$-charges which are integer multiples 
of~$X_{\mathfrak{A}}$: In such a case, the operator in Eq.~(\ref{aHoch})
produces  
$
{\langle \mathfrak{A} \rangle }^{-({n_{\varPhi_1}\!\cdot X_{\varPhi_1}+\,n_{\varPhi_2}\!\cdot X_{\varPhi_2}+...+\,n_{\varPhi_{k_{\mathrm{max}}}}\!\cdot
X_{\varPhi_{k_{\mathrm{max}}}}})/{X_{\mathfrak{A}}}} \cdot T$.  
On the other hand, terms which are undesired (like {\it e.g.} baryon number
violating operators) might be assigned an overall $X$-charge which is not an
integer multiple of $X_{\mathfrak{A}}$ so that the exponent of $\mathfrak{A}$
is fractional and the whole term thus forbidden. Therefore not all
SM-invariant terms are necessarily generated, because the corresponding
``mother terms'' might be forbidden due to the CDP's persistent constraints.
 \emph{These omissions are  what one calls ``forbidden due to a DGS'', the DGS being the remnant/residual/left-over of a spontaneously broken local 
gauge symmetry.} If a super- or K\"ahler potential term is forbidden, then the $|X_\mathfrak{A}|^{th}$ power of this term is allowed for sure. This reasoning is precisely the same as the one which we reviewed in the beginning of Sect.~\ref{discretesect}, see also Eq.~(\ref{ztotal}). \\
To parameterize the possible deviation of  $n_{\varPhi_1}\!\cdot X_{\varPhi_1}+\,n_{\varPhi_2}\!\cdot X_{\varPhi_2}+...+\,n_{\varPhi_{k_{\mathrm{max}}}}\!\cdot
X_{\varPhi_{k_{\mathrm{max}}}}$ from being an integer multiple of
$|X_\mathfrak{A}|$, one introduces the following  decomposition of the
$X$-charges 
\beq\label{decomp}
X_{\varPhi_j}=m_{\varPhi_j}\cdot |X_\mathfrak{A}|+ z_{\varPhi_j}\ .
\eeq
$m_{\varPhi_j}$ and the discrete charge $z_{\varPhi_j}$ are both
integer, the latter being  restricted to $\{0,1,...,|X_\mathfrak{A}|-1\}$. So
if the  sum of the $z_{...}$ of several superfields does not produce
an integer multiple of $X_\mathfrak{A}$, the corresponding term is 
not allowed; we have a $\mathds{Z}_{|X_\mathfrak{A}|}$-symmetry. In the
following we are going to work with the standard notation: 
$$
|X_\mathfrak{A}|\equiv N \: .
$$
The $N$ above however might not yet be the one showing up 
in ``$\mathds{Z}_N$'':   Suppose that $N=24$, then the superfields suggest a
$\mathds{Z}_{24}$-symmetry. But it might well be that for all SM gauge
invariant operators the overall discrete charges are even, so that
rescaling at the operator level effectively yields a $\mathds{Z}_{12}$-symmetry.

\item \label{texture} We demand that the $X$-charges of the superpotential terms
  $Q^iH_d\overline{D^j}$  and $Q^iH_u\overline{U^j}$ ($i,j\in\{1,2,3\}$) are
  integer multiples  of  $N$. Otherwise the corresponding Yukawa coupling
  constants would contain zero-entries due to the CDP, which would translate
  to unobserved zero-entries in the Cabibbo-Kobayashi-Maskawa (CKM) matrix. So
  we find that the discrete   charges of the quarks have to be
  generation-independent, although the   original $X$-charges might well be
  generation-dependent:  $m_{Q^i}\neq m_{Q^j}$ but  $z_{Q^i}= z_{Q^j} \equiv
  z_Q$, see Eq.~(\ref{decomp}). In other words, discrete quark charges are
  family-universal.    

\item For simplicity, we demand the same for the leptons.\footnote{If one
    relies on Dirac neutrinos or a see-saw, the same arguments as in
    Item~\ref{texture} apply, with the Maki-Nakagawa-Sakata matrix \cite{Maki:1962mu}
    replacing the CKM matrix. One way to avoid this conclusion is the
    generation of (Majorana) neutrino masses from  loop-effects, see
    Ref.~\cite{Kapetanakis:1992jj}.} 
With only generation-independent discrete charges and the requirement that the
three SSM Yukawa couplings are allowed by the discrete symmetry, {\it i.e.}
\beqn
z_Q+z_{H_d}+z_{\overline{D}}&=&0~\mathrm{mod}~N \ ,\nonumber\\
z_Q+z_{H_u}+z_{\overline{U}}&=&0~\mathrm{mod}~N \ , \label{dreiyuks}\\
z_L+z_{H_d}+z_{\overline{E}}&=&0~\mathrm{mod}~N \ ,\nonumber
\eeqn
the total discrete charge of any gauge-invariant term in the
$\mathfrak{A}+$SSM sector can be expressed as, see also
Refs.~\cite{Dreiner:2005rd,Dreiner:2006xw,Dreiner:2007vp},
\beqn\label{ztotal}
z_{\mathrm{total}}& =& 
    (z_{\overline{U}}+z_{\overline{D}}+z_{\overline{D}})\cdot\mathbb{Z}
     +  (z_{Q}+z_{L}+ z_{\overline{D}})\cdot\mathbb{Z} \nonumber \\ 
&+&  (z_{H_d}+z_{H_u}) \cdot \mathbb{Z}+ N\cdot \mathbb{Z}\ ,
\eeqn
with $\mathbb{Z}$ representing an integer number. This result motivates the
classification of the $\mathds Z_N$-symmetries in Table~\ref{asasassa}.

\item \label{rmnp} We now add the three right-handed neutrinos
$\overline{\mathcal N}$ to the theory, additionally requiring that the Yukawa
terms $LH_u\overline{\mathcal N}$ are allowed by the DGS,
\beqn
z_L+z_{H_u}+z_{\overline{\mathcal{N}}}&=&0~\mathrm{mod}~N \ . \label{einyuk}
\eeqn
Solving the four equations of Eqs.~(\ref{dreiyuks},\ref{einyuk}) with eight
unknowns, we can express the $z_{...}$ in terms of the four parameters
$m,n,p,r\in\{0,1,...,N-1\}$ (so $ QH_d \overline{D}$,  $QH_u \overline{U}$, 
$LH_d \overline{E}$, and $LH_u\overline{\mathcal{N}}$ are required, but 
not $H_dH_u$): 
\begin{eqnarray}
z_Q\!\!\!&=&\!\!\! r\ , \quad \qquad \qquad \qquad
z_{\overline{D}}=m-n+2r \ , \quad \;\; z_{\overline{U}}=-m-4r        \ ,
\notag \\
\quad z_L\!\!\!&=&\!\!\!-n-p-3r\ , \quad \quad\, z_{\overline{E}}=m+p+6r\ , \quad ~ \, z_{\overline{\mathcal{N}}}=-m+n+p \ ,  \notag \\
z_{H_d}\!\!\!&=&\!\!\!
-m+n-3r\ , \quad \,\, ~z_{H_u}=m+3r\ .\label{withr}
\end{eqnarray}
The coefficient of $r$ is proportional to the hypercharge
of the corresponding particle (see Table~\ref{commonDS}); hence $r$
is the discrete version of the $Y$-shift-parameter $\alpha$ in 
Item~\ref{yshift}. Choosing $r=0$, we recover the same parameterization of
discrete symmetries as in Ref.~\cite{Ibanez:1991pr}, here generalized to
include the right-handed neutrinos. 

\item \label{greencon} The $X$-charges decompose according to
  Eq.~(\ref{decomp}). Using Eq.~(\ref{withr}), we can rewrite the
  Green-Schwarz anomaly cancellation conditions in terms of the discrete 
parameters
  $m,n,p,r$. For instance, for the anomaly $\mathcal A_{CCX}$ we obtain, see
  Eqs.~(\ref{c-anom},\ref{x-eq}),
\beq\label{EQ18}
\frac{1}{2k_C}\left[{-N_f\cdot n}+\sum_{i=1}^{N_f}(2
  m_{Q^i}+m_{\overline{D^i}}+ m_{\overline{U^i}} )\cdot N
  +2\cdot\mathcal A_{CCX}^\mathrm{beyond~MSSM}\right] ~=~ 2\pi^2X_S \ ,
\eeq
with $N_f$ denoting the number of generations. This, however, does not specify
everything, since we have not yet dealt with beyond-MSSM matter.

\item \label{cwy} In the following, we list our assumptions about SM-charged matter which
  is not part of the MSSM: 
\begin{itemize}
\item $C$-charged matter: There may be no massless colored particles, as these
would have been seen already by experiment. What can in principle occur is
colored matter in vectorlike pairs which is too heavy to have been detected so
far. After $U(1)_X$ breaking, the corresponding mass terms must therefore be
$\mathds{Z}_N$-invariant.
\item $W$-charged matter: As for colored particles.  
\item $Y$-charged matter: We distinguish the following mutually independent cases, elucidated below
\begin{center}
\begin{tabular}{||l||l|l||}
\hline\hline 
 &  \begin{tabular}{c}$Y$-charge is normal\\ or large compared \\to SM $Y$-charges. \end{tabular}& \begin{tabular}{c}$Y$-charge is tiny \\ compared to \\ SM $Y$-charges. \end{tabular} \\
\hline\hline
Beyond-SM matter is heavy. & {(a)} ~~~o.k. & {(b)}  \begin{tabular}{c}  renders $\mathcal{A}_{YYX}$ and\\  $\mathcal{A}_{YXX}$ useless\end{tabular} \\
\hline
\begin{tabular}{c}Beyond-SM matter is light \\but  not 
massless. \end{tabular} & {(c)}   \begin{tabular}{c} not observed \\ by experiment \end{tabular} &  {(d)} \begin{tabular}{c}  renders $\mathcal{A}_{YYX}$ and\\  $\mathcal{A}_{YXX}$ useless\end{tabular} \\
\hline
Beyond-SM matter is massless. & {(e)} \begin{tabular}{c} not observed \\ by experiment \end{tabular} & {(f)}  \begin{tabular}{c}  renders $\mathcal{A}_{YYX}$, $\mathcal{A}_{YXX}$\\ and $\mathcal{A}_{GGX}$ useless\end{tabular}\\
\hline\hline
\end{tabular}
\end{center}
\newpage
\begin{enumerate}
\item[(a)]Just like before, heavy particles with reasonable $Y$-charges are
  acceptable.
\item[(b,d)] Heavy or light-but-not-massless particles with tiny $Y$-charges
 cannot be ruled out. The presence of such particles spoils the predictability
 of $\mathcal{A}_{YYX}$ and $\mathcal{A}_{YXX}$, which is why we shall not use
 these two constraints; for more details see Sect.~4 of
 Ref.~\cite{Dreiner:2005rd}.   
\item[(c,e)] There may be no light or even massless particles with a 
reasonable, {\it i.e.} not too small, hypercharge, as these would have been 
seen already by experiment.

\item[(f)]In principle, one could also have massless particles with tiny (experimentally yet undetectable) hypercharges. Then, however, a systematic analysis of the discrete anomaly condition would not be possible. Hence, we demand such particles to be absent.

\end{enumerate}
\end{itemize}
With these assumptions Eq.~(\ref{EQ18}) reads
\beq\label{holland}
{-N_f\cdot n}+N\cdot \mathbb{Z}~=~ 4~\pi^2~X_S~ k_C \ ,
\eeq
$\mathbb{Z}$ symbolizing an integer number. A similar relation is obtained 
for $\mathcal A_{WWX}$. 

\item \label{katze} We demand unification of the three MSSM gauge coupling
  constants. That is, adopting the hypercharge normalization
  $Y_L=\frac{1}{2}$, we require that the Ka\v{c}-Moody levels are related by
  $k_C=k_W=\frac{3}{5} k_Y$, see Eq.~(\ref{g^2_mal_k}).   

\item \label{it21} We demand the $U(1)_X$ to be anomaly-free, \emph{i.e.} $X_S=0$. (This makes Item~\ref{katze} superfluous.) 

\item \label{miststueck} The only massless $\Omega$-type particles (see
  Item~\ref{AundO}) we 
shall admit are right-handed neutrino superfields~$\overline{\mathcal{N}}$,
\emph{i.e.} particles whose discrete charge is such that their trilinear
coupling  to $LH_u$ is allowed, {\it cf.} Eq.~(\ref{einyuk}). Massive
$\Omega$s can be assumed as well without spoiling the analysis in
Sect.~\ref{mainpart}. Other types of particles are classified in
Appendix~\ref{singclas}. In the language of Appendix~\ref{singclas}, we shall
deal with ``Case~3'', which has the term $LH_uLH_u$ not allowed, thus
we will not have to deal with pseudo-Dirac neutrinos. 
Having constrained the SM-singlet particle content, the calculation of
the gravitational anomaly $\mathcal A_{GGX}$ becomes feasible, as well. 
Now, Eq.~(\ref{holland}) and Item~\ref{it21} together with the equivalent relations for $\mathcal
A_{WWX}$ and  $\mathcal A_{GGX}$ lead to the starting point of our
investigation in the next section:
\beqn\label{didi_dirac}
 -N_f  \cdot n + N \cdot \mathbb{Z} &=&0\ , \\
 -N_f\cdot  (n+p) + N_H\cdot n + N \cdot
\mathbb{Z} &=&0 \ , \\
 -N_f\cdot  (5n +p-m-\zeta_{\overline{\mathcal{N}}}) + 2N_H \cdot n + N \cdot \mathbb{Z} +
\eta \; \frac{N}{2} \cdot \mathbb{Z} &=&0 \ .~~~~\label{dodo_dirac}
\eeqn
$N_H$ is the number of pairs of Higgs doublets. $\eta=0,1$ for
$N=\mathrm{odd,even}$; furthermore, 
$\zeta_{\overline{\mathcal{N}}}=0$ in a theory without light right-handed
neutrinos and $\zeta_{\overline{\mathcal{N}}}=-m+n+p$ if there are 
$N_f$ generations of~$\overline{\mathcal{N}}$.
Note that the $r$-dependence drops out since the linear anomalies are
invariants under $Y$-shifts, see Item~\ref{yshift}.

 \item Finally we integrate out heavy degrees of freedom, including the heavy
$U(1)_X$ gauge boson. This might cause a rescaling of the discrete charges,
for the MSSM sector could have a discrete symmetry which is a subgroup of the 
overall $\mathds{Z}_N$-symmetry. Consider again the example of a
$\mathds{Z}_{24}$. Suppose that all MSSM superfields have even discrete
charges, but some heavy particles have $z=1$. Then, the $\mathds{Z}_{24}$
cannot be rescaled to a $\mathds{Z}_{12}$ like in the example at the end of
Item~\ref{VI}. However, after the energies have dropped below the masses of
the  $z=1$ heavy matter, one can integrate it out, and a rescaling (now only
within the MSSM sector) becomes possible.

\end{enumerate}

%%%%%%%%%%%%%%%%%%%%%%%

%%%%%%%%%%%%%%%%%%%%%%%

%%%%%%%%%%%%%%%%%%%%%%%

\section{\label{mainpart}Anomaly-free Dirac-DGSs}
\cleqn

Compared with Refs.~\cite{Ibanez:1991pr,Dreiner:2005rd}, we have added three
right-handed neutrinos $\overline{\mathcal{N}}$ to the {\it light} particle
content. Analogously to \cite{Dreiner:2005rd} we now discuss the
resulting discrete anomaly conditions, {\it i.e.}
Eqs.~(\ref{didi_dirac}-\ref{dodo_dirac}). 
Note that in a scenario with light right-handed neutrinos the parameter $m$
remains unconstrained; so regardless of what the values for $N_f$ and  
$N_H$ are, it can take all $N$ values $m=0,1,...,N-1$. Restricting 
ourselves to $N_f=3$ and $N_H=1$, we get
\beqn
-3 n & = & N \cdot \mathbb{Z}\ ,\label{38}\\
-2 n  - 3  p & = &  N \cdot \mathbb{Z}\ ,\label{39}\\
3 m -13n-3 p+3\zeta_{\overline{\mathcal{N}}} & = &   N \cdot \mathbb{Z} +
\eta \; \frac{N}{2} \cdot \mathbb{Z}\ ,\label{310}
\eeqn  
which can be linearly combined to give
$3 n  =  N \cdot \mathbb{Z}$, 
$3  p - n =  N \cdot \mathbb{Z}$ and [$(\ref{310})-2\times(\ref{39})-3\times(\ref{38})$]
$3 (m+p+\zeta_{\overline{\mathcal{N}}}) =    N \cdot \mathbb{Z} +
\eta \; \frac{N}{2} \cdot \mathbb{Z}$.
If  $\zeta_{\overline{\mathcal{N}}}=0$ we recover Eqs.~(2.21--2.23) of
Ref.~\cite{Dreiner:2005rd}; plugging in
$\zeta_{\overline{\mathcal{N}}}=-m+n+p$ , {\it i.e.} considering the case with
three $\overline{\mathcal{N}}$ (which from the viewpoint of the discrete
anomaly conditions could have Majorana 
mass terms if $\zeta_{\overline{\mathcal{N}}}=0,{N}/{2}$), yields 
\beqn
 3 n & = & N \cdot \mathbb{Z}\ , \label{eq25}\\
3 p-n  & = &  N \cdot \mathbb{Z}\ ,\label{eq26}\\
6p + 3n & = &   N \cdot \mathbb{Z} + \eta \; \frac{N}{2} \cdot \mathbb{Z}\ .\label{eq27}
\eeqn
The calculation of $3 \times (\ref{eq25}) + 4 \times (\ref{eq26}) -2 \times
(\ref{eq27})$ leads to the condition $-n  =  N \cdot \mathbb{Z}$ which reveals
that $n=0$, thus rendering Eq.~(\ref{eq25}) trivial. Interestingly enough,
this is exactly the condition for having the bilinear term  $H_d H_u$ allowed
by the discrete symmetry, since $z_{H_dH_u}=n$, see Eq.~(\ref{withr}).
\emph{So without demanding it, the $\mu$-term emerges automatically due to
anomaly considerations,} unlike in
Refs.~\cite{Ibanez:1991pr,Dreiner:2005rd}.  From Eq.~(\ref{eq26}) we now
obtain   
\beqn
3p & = & N \cdot \mathbb{Z}\ .\label{pconstraint}
\eeqn
Only in those cases where $N$ is a multiple of three, $p$ can take a
non-trivial value. However, there exist non-trivial DGSs also with $p=0$,
taking {\it   e.g.}  $m=N/2$ gives~$M_p$. 

With right-handed neutrinos, all anomaly-free DGSs can now be classified
by the set of integers $(N;m,n,p) = (N;m,0,p)$, with the constraint of
Eq.~(\ref{pconstraint}). In contrast to Ref.~\cite{Dreiner:2005rd}, a
Majorana mass term $\overline{\mathcal{N}}\overline{\mathcal{N}}$ is not
imposed here. As this term has discrete charge $2(p-m)$, it is allowed only
if either $\left( p=0 \, \wedge \, m=\frac{N}{2}\right)$, $\left(p=\frac{N}{3}
  \, \wedge \, 
m=\frac{N}{3},\frac{5N}{6}\right)$ or $\left(p=\frac{2N}{3} \, \wedge \,
m=\frac{2N}{3},\frac{N}{6}\right)$; of course, $N$ must be divisible by 
2~and/or~3.
The classification of the DGSs in terms of the values of $N$ is shown in
Table~\ref{table4}. 
\begin{table}[t]
\begin{center}
\hspace{0mm}\footnotesize\begin{tabular}{||c||c|c||}
\hline\hline $\phantom{\Big|}$ & $ (2|N)$ &  $\neg(2|N)$\\\hline\hline
  $ (3|N) $    &  
\begin{tabular}{c} ~\\$N=6,12,18,24,%30,36,42,48,54,
...$ \\~\\
$p=0$ \begin{tabular}{l} $m=0$ $\Rightarrow$ trivial\\
  $m=N/2$ $\Rightarrow$ $M_p$ and $\exists\overline{\mathcal{N}}\overline{\mathcal{N}}$\\
  else $\Rightarrow$ ``new'' DGS ($\not\!\exists\overline{\mathcal{NN}}$)\end{tabular}\\~\\\hline
 $p=N/3$ 
\begin{tabular}{l} ~\\$m=N/3$ $\Rightarrow$  $B_3$ and $\exists\overline{\mathcal{N}}\overline{\mathcal{N}}$\\
  $m=5N/6$ $\Rightarrow$ $P_6$ and $\exists\overline{\mathcal{N}}\overline{\mathcal{N}}$\\
  else $\Rightarrow$ ``new'' DGS ($\not\!\exists\overline{\mathcal{N}\mathcal{N}}$)\\~\end{tabular}
\\\hline
$p=2N/3$ \begin{tabular}{l}~\\ $m=2N/3$ $\Rightarrow$  $B_3$ and $\exists\overline{\mathcal{N}}\overline{\mathcal{N}}$\\
  $m=N/6$ $\Rightarrow$ $P_6$ and $\exists\overline{\mathcal{N}}\overline{\mathcal{N}}$\\
  else $\Rightarrow$ ``new'' DGS ($\not\!\exists\overline{\mathcal{N}\mathcal{N}}$)\end{tabular}\\~
\end{tabular}        &   
\begin{tabular}{c} $N=3,9,15,21,27,%33,39,45,51,
...$ 
\\~\\$p=0$ \begin{tabular}{l} $m=0$ $\Rightarrow$ trivial\\
  else $\Rightarrow$ ``new'' DGS ($\not\!\exists\overline{\mathcal{N}\mathcal{N}}$)\end{tabular}\\~\\\hline $p=N/3$ 
\begin{tabular}{l} ~\\$m=N/3$ $\Rightarrow$  $B_3$ and $\exists\overline{\mathcal{N}}\overline{\mathcal{N}}$\\
  else $\Rightarrow$ ``new'' DGS ($\not\!\exists\overline{\mathcal{N}\mathcal{N}}$)\\~\end{tabular}
\\\hline
$p=2N/3$ \begin{tabular}{l} ~\\$m=2N/3$ $\Rightarrow$  $B_3$ and $\exists\overline{\mathcal{N}}\overline{\mathcal{N}}$\\
  else $\Rightarrow$ ``new'' DGS ($\not\!\exists\overline{\mathcal{N}\mathcal{N}}$)\\~\end{tabular}
\end{tabular}          \\\hline
   $\neg(3|N)$    &  
\begin{tabular}{c} ~\\$N=2,4,8,10,14,16,20,22,26,...$   \\
$p=0$ \begin{tabular}{l} ~\\$m=0$ $\Rightarrow$ trivial\\
  $m=N/2$ $\Rightarrow$ $M_p$ and $\exists\overline{\mathcal{N}}\overline{\mathcal{N}}$\\
  else $\Rightarrow$ ``new'' DGS ($\not\!\exists\overline{\mathcal{N}\mathcal{N}}$)\\~\end{tabular} \end{tabular}    & 
\begin{tabular}{c} $N=5,7,11,13,17,19,23,25,...$ \\~\\
$p=0$~\begin{tabular}{l}  $m=0$ $\Rightarrow$ trivial\\
  else $\Rightarrow$ ``new'' DGS ($\not\!\exists\overline{\mathcal{N}\mathcal{N}}$)\end{tabular} 
\end{tabular}          \\\hline\hline
\end{tabular}
\end{center}
\caption{\label{table4}Classifying the anomaly-free DGSs with 
  right-handed neutrinos in terms of the value for $N$. $\neg(2|N)$ and
  $\neg(3|N)$ denotes that $N$ is not an integer multiple of 2 and 3,
  respectively. Note that the treatment in this section is more  general 
  than the one in Ref.~\cite{Dreiner:2005rd}; ``without Dirac neutrinos'' 
  is so-to-speak a special case of ``with Dirac neutrinos'', see also 
   Table~\ref{om(eg)a} (no Dirac neutrinos means ``not Case~3'', so Cases~1 and~2 remain, both with and without right-handed neutrinos $\mathcal{N}_{\mathrm{Maj}}$).}
\end{table}
The cases allowing for the Majorana mass term
$\overline{\mathcal{N}}\overline{\mathcal{N}}$ are listed explicitly and
correspond to $M_p$, $B_3$ and $P_6$ only.  
In order to comply with our requirement of having pure Dirac neutrinos, see
Item~\ref{miststueck}, we discard these solutions of the anomaly conditions. 
All other cases, however, yield \emph{new} anomaly-free DGSs, which we will call \emph{Dirac-DGSs}.  The $\mathds{Z}_N$-symmetries up to $N=6$ 
are given in 
Table~\ref{table5}, just to list a few. 
\begin{table}[t]
{
\begin{center}\scriptsize\begin{tabular}{||c|c|c|c|c||}
\hline\hline
$N$ & $m$ & $n$ & $p$ & DGS \\\hline\hline 
2 & 1& 0&  0& $M_p$ (not a Dirac-DGS)\\\hline
3 & 0& 0&  0&  trivial\\
  & 0& 0&  1&  "new"\\
  & 0& 0&  2&  same as 2nd\\
  & 1& 0&  0&  "new"\\
  & 1& 0&  1&   $B_3$ (not a Dirac-DGS)\\
  & 1& 0&  2&  "new"\\
  & 2& 0&  0&  same as 4th\\
  & 2& 0&  1&  same as 6th\\
  & 2& 0&  2&  same as 5th\\\hline
4 & 0 & 0&  0 & trivial \\
  & 1 & 0&  0 & "new"\\
  & 2 & 0&  0 & $M_p$ (not a Dirac-DGS)\\
  & 3 & 0&  0 & same as 2nd\\\hline

5 & 0 & 0&  0 & trivial \\
  & 1 & 0&  0 & "new"\\
  & 2 & 0&  0 & "new"\\
  & 3 & 0&  0 & same as 3rd\\
  & 4 & 0&  0 & same as 2nd\\\hline
6  & 0  & 0&  0 &   trivial  \\
  & 0  & 0&  2 &    same as (3;0,0,1) \\
  & 0  & 0&  4 &    same as 2nd \\
  & 1  & 0&  0 &    "new" \\
  & 1  & 0&  2 &    "new" \\
  & 1  & 0&  4 &     same as 17th\\
  & 2  & 0&  0 &     same as (3;1,0,0)\\
  & 2  & 0&  2 &     $B_3$ (not a Dirac-DGS)\\
  & 2  & 0&  4 &     same as (3;1,0,2)\\
  & 3  & 0&  0 &     $M_p$ (not a Dirac-DGS)\\
  & 3  & 0&  2 &     "new"\\
  & 3  & 0&  4 &     same as 11th\\
  & 4  & 0&  0 &     same as 7th\\
  & 4  & 0&  2 &     same as 9th\\
  & 4  & 0&  4  &     same as 8th\\
  & 5  & 0&  0 &     same as 4th\\
  & 5  & 0&  2 &     $P_6$  (not a Dirac-DGS)\\
  & 5  & 0&  4 &     same as 5th\\\hline\hline
\end{tabular}\end{center}
\caption{\label{table5}The easiest $\mathds{Z}_N$-symmetries. The comment ``same as $x$th''
  means that the symmetry is equivalent to the one in the $x$th line
  of the symmetries with identical $N$.}}
\end{table} 
Thus, excluding \emph{1.)} $\{M_p,B_3,P_6\}$, \emph{2.)} rescalings like $(6;0,0,2 )=(3; 0,0,1)$ as well as \emph{3.)} double
counting like $( 3; 0,0 ,1 )=( 3; 0,0 ,2 )$  [$(N;m,0,p)$, $(N;N-m,0,N-p)$ and $(a\cdot N;a\cdot m,0,a \cdot p)$, with $a$ being a positive integer, give the same DGS, for the latter see Item~\ref{VI}], we have many Dirac-DGSs,\footnote{Having taken rescaling already into account, why is the
  number of DGSs in the case without Dirac neutrinos three and in the
  case with Dirac neutrinos $\infty$?  With Majorana neutrinos, the
  possibility of rescaling the 
  discrete charges leads to a finite number of distinct DGSs
  \cite{Dreiner:2005rd}. But allowing for Dirac neutrinos, the parameter $m$
  is not constrained at all, therefore the choice of $N$ being an arbitrary
  prime number always leads to non-trivial Dirac-DGSs.}   
also with   $N\leq6$:\footnote{Like the numerical syllables in
  \emph{tri}ality 
  and \emph{hexa}lity, we shall stick to   Greek rather
  than Latin. Otherwise we would have \emph{e.g.} tertiality, quartality,
  quintality, sextality and septality.} 
\begin{itemize}
\item three trialities: $( 3; 0,0 ,1 )$,  $( 3; 1,0 ,0 )$, $( 3; 1,0 ,2 )$\ ,
\item one tetrality: $( 4; 1,0 ,0 )$\ ,
\item two pentalities: $( 5; 1,0 ,0 )$\ , 
$( 5; 2,0 ,0 )$\ ,
\item three hexalities: $( 6; 1,0 ,0 )$, $( 6; 1,0 ,2 )$, $( 6; 3,0 ,2 )$\ .
\end{itemize}
Beyond Table~\ref{table5} we easily also find
\begin{itemize}
\item three heptalities: $(7; 1, 0, 0)$, $(7; 2, 0, 0)$, $(7; 3, 0, 0)$\ , 
\item two octalities: $(8; 1, 0, 0)$, $(8; 3, 0, 0)$\ ,
\item continuing to, say,  $N=14$, there are (all distinct) nine 9-alities, two 10-alities, five
11-alities, eight 12-alities, six 13-alities, three 14-alities, see also the Table in Appendix~\ref{gutapp}.

\end{itemize}

Before  the discussion of the physical implications of the
Dirac-DGSs, some comments concerning the purely Abelian anomaly conditions are
in order. 

\begin{itemize}
\item  As observed in Ref.~\cite{Dreiner:2005rd} and Item~\ref{cwy}, the anomaly coefficients
$\mathcal{A}_{YYX}$ and $\mathcal{A}_{YXX}$ do not pose useful constraints on
the DGSs because the hypercharges of heavy Dirac particles could 
be fractional; this statement holds true for Dirac-DGSs as well. 

\item On the other
hand, in Ref.~\cite{Dreiner:2005rd} $\mathcal{A}_{XXX}$ contained information
about whether or not fractionally $X$-charged exotic matter has to be assumed
for a given DGS. This is {\it not} the case for Dirac-DGSs as we will sketch
in the  following. It was shown in Ref.~\cite{Dreiner:2005rd} that the cubic
anomaly condition  $\mathcal{A}_{XXX} =0$ can be written as $
\sum_i z_i^3 ~=~\mathrm{RHS}
$,
with the $z_i$ denoting the discrete charges of the particles in the $\overline{\mathcal N}$+SSM sector. The RHS
can take on only certain values depending on $N$, {\it cf.}  \cite{Dreiner:2005rd}'s Eqs.~(A.3,A.4):
\begin{center}
\begin{tabular}{||r||c|c|c||}
\hline\hline
RHS~$\phantom{\Big|}$ & $\neg (2|N)$ &  $(2|N) ~ \wedge ~ \neg (4|N) $ & $(4|N)$ 
\\\hline\hline
$\neg (3|N)\phantom{\Big|}$ &  $N \cdot \mathbb{Z}$ &  $\frac{N}{2} \cdot
\mathbb{Z}$ &  $N \cdot \mathbb{Z}$ 
\\ \hline 
$(3|N)\phantom{\Big|}$ &    $3 N \cdot \mathbb{Z}$ &  $3 \frac{N}{2} \cdot
\mathbb{Z}$ &    $3 N \cdot \mathbb{Z}$  \\\hline\hline 
\end{tabular}
\end{center}
These possible values for the RHS must be compared to the sum over
$z_i^3$, which we can express in terms of the parameters $(m,n,p)$. In
contrast to~\cite{Dreiner:2005rd}, we now have to include the three
right-handed neutrinos with discrete charge $(p+n-m)$; 
this simplifies the resulting expression, see Eq.~(A.1) 
of~\cite{Dreiner:2005rd},  considerably. Inserting $n=0$, a necessity for
all Dirac-DGSs, we get$
\sum_i z_i^3 ~=~18 \, m^2 \,p
$,
which, due to Eq.~(\ref{pconstraint}), is always an integer multiple of
$6m^2\,N$. The RHS can match this value for all possible values of
$N$. One therefore does not have to rely on fractionally $X$-charged heavy
particles in order to meet the cubic anomaly condition. In this respect, $\mathcal{A}_{XXX}$ does not constrain the Dirac-DGSs.
\end{itemize}

%%%%%%%%%%%%%%%%%%%%%%%%%%%%%%

%%%%%%%%%%%%%%%%%%%%%%%%%%%%%%

%%%%%%%%%%%%%%%%%%%%%%%%%%%%%%

\section{\label{physicssetc}The physics of Dirac-DGSs}
\cleqn
In order to discuss the physical implications of the  Dirac-DGSs, we
investigate which lepton and/or baryon number violating operators are allowed
for these new symmetries. As mentioned in Section~\ref{discretesect}, many
of these operators come together with other operators if one assumes the
presence of the MSSM superpotential terms, see
Eq.~(\ref{BVLVoperators}). Even though the $\mu$-term is initially not
required, it arises automatically for Dirac-DGSs 
due to anomaly considerations. Therefore, the classification of the lepton
and/or baryon number violating operators up to dimension five given in
Eq.~(\ref{BVLVoperators}) applies to the Dirac-DGSs as well. 

However, in the
Dirac case, there is a new particle, the right-handed 
neutrino~$\overline{\mathcal{N}}$ with $n_{\mathrm{lepton}}=-1$, which leads
to additional SM invariant terms. We have to determine these new operators and
group them together depending on their discrete charges: If, under a specific
DGS $(N;m,0,p)$, one term has for example discrete charge $p$ and another has
charge $-p$, then both operators are simultaneously forbidden ($p\neq 0$) or
allowed ($p=0$). The resulting sets of operators up to dimension five are, see
Appendix~\ref{endlichregen!},
\begin{eqnarray}
LH_u\overline{\mathcal{N}} &\Longleftrightarrow & \Big\{
LH_u\overline{\mathcal{N}}, \,   
LL\overline{E}\overline{\mathcal{N}} , \,
QL\overline{D}\overline{\mathcal{N}} \Big\}\ , \nonumber \\
QL\overline{D} &\Longleftrightarrow & \Big\{ 
\overline{\mathcal{N}},\, 
H_dH_u\overline{\mathcal{N}},\,
QH_d\overline{D}\overline{\mathcal{N}},\,
QH_u\overline{U}\overline{\mathcal{N}},\,
LH_d\overline{E}\overline{\mathcal{N}},\,
LH_u\overline{\mathcal{N}}\overline{\mathcal{N}} \Big\}\ , \nonumber \\
LH_uLH_u &\Longleftrightarrow & \Big\{ 
\overline{\mathcal{N}}\overline{\mathcal{N}},\,
H_dH_u\overline{\mathcal{N}}\overline{\mathcal{N}} \Big\}\ , \nonumber\\
QQQL &\Longleftrightarrow & \Big\{
\overline{U}\overline{D}\overline{D}\overline{\mathcal{N}}\Big\}\ ,
\label{BVLVwithN}\\ 
\overline{\mathcal{N}}\overline{\mathcal{N}}\overline{\mathcal{N}} 
 &\Longleftrightarrow & \Big\{
\overline{\mathcal{N}}\overline{\mathcal{N}}\overline{\mathcal{N}} \Big\}\ ,
\nonumber\\
\overline{\mathcal{N}}\overline{\mathcal{N}}\overline{\mathcal{N}}\overline{\mathcal{N}} 
&\Longleftrightarrow & \Big\{
\overline{\mathcal{N}}\overline{\mathcal{N}}\overline{\mathcal{N}}\overline{\mathcal{N}}\Big\}
\ . \nonumber
\end{eqnarray}
The second, third and fourth line generalize Lines~1, 4 and~3 of Eq.~(\ref{BVLVoperators}). The terms in the first line do {\it not} violate lepton or baryon number and
are, by definition [see Eq.~(\ref{einyuk})], always allowed by
Dirac-DGSs. We therefore focus on the remaining sets.
Taking into account also the operators of Eq.~(\ref{BVLVoperators}), we 
obtain six sets of $n_{\mathrm{lepton}}$- and/or
$n_{\mathrm{quark}}$-violating operators, which can be represented by the
terms  
\begin{equation}
QL\overline{D} \, ,~ 
LH_uLH_u \,,~
\overline{U}\overline{D}\overline{D} \,,~
QQQL \,,~
\overline{\mathcal{N}}\overline{\mathcal{N}}\overline{\mathcal{N}} \,,~
\overline{\mathcal{N}}\overline{\mathcal{N}}\overline{\mathcal{N}}\overline{\mathcal{N}}
\ . \label{physops}
\end{equation}
Since our focus is to classify the Dirac-DGSs, {\it i.e.} those which forbid
the Majorana mass term $\overline{\mathcal{N}}\overline{\mathcal{N}}$,  the $QL\overline{D}$-set ($\ni \overline{\mathcal{N}}$) and the 
$LH_uLH_u$-set ($ \ni \overline{\mathcal{N}}\overline{\mathcal{N}}$) are
never allowed by Dirac-DGSs. \emph{Comparing with Table~\ref{asasassa} shows that
Dirac-DGSs can never be {baryon} $N$-alities, but only {matter}  or
{lepton} $N$-alities.}\footnote{It is easy to check which type of $N$-ality one
  obtains from a given parameter set $(N;m,n,p)$. Using Eq.~(\ref{withr}), the
  discrete   charges of $QL\overline{D}$ and  $\overline{UDD}$ are given as
  $z_{QL\overline{D}} = m-2n-p$ and $z_{\overline{UDD}}= m-2n$,
  respectively. With Table~\ref{asasassa} we find that {baryon} $N$-alities
  require   $[\,z_{\overline{UDD}}\neq 0 
  \,\wedge\, z_{QL\overline{D}}= 0\,]$, {lepton} $N$-alities must have
  $[\,z_{\overline{UDD}}= 0 \,\wedge\, z_{QL\overline{D}}\neq 0\,]$, and
  {matter} $N$-alities need $[\,z_{\overline{UDD}}\neq 0\,\wedge\,
  z_{QL\overline{D}}\neq 0\,]$.}   
Let us therefore discuss the remaining four sets in turn.
\begin{itemize}
\item \underline{$\overline{U}\overline{D}\overline{D}$:} Under the general
  Dirac-DGS  $(N;m,0,p)$, the discrete charge of these operators is $\pm
  m$. They are thus present in theories where the DGS has $m=0$. With
  Eq.~(\ref{pconstraint}) these are $(N;0,0,N\cdot \mathbb Z /3)$, leading to
  lepton triality $(3;0,0,1)$ as the only possibility after rescaling.
  All other Dirac-DGSs forbid $\overline{U}\overline{D}\overline{D}$ and its
  accompanying operators, {\it cf.} Eq.~(\ref{BVLVoperators}); {\it they are
  therefore all {matter} $N$-alities}. 
\item \underline{$QQQL$:} The discrete charge for this set of operators is given by $\mp
  p$. They are therefore present in {\it all}$\;\,\mathds{Z}_N$-symmetries with
  $N\neq 3\cdot \mathbb{Z}$ [see Eq.~(\ref{pconstraint})] as well as in those
  symmetries with $N=3\cdot \mathbb{Z}$ and $p=0$.
\item  \underline{$\overline{\mathcal{N}}\overline{\mathcal{N}}\overline{\mathcal{N}}$:} 
 This cubic operator carries charge $3(p-m)$. Due to Eq.~(\ref{pconstraint}),
 this is equivalent to the discrete (mod~$N$) charge $3m$. Hence, 
 $\overline{\mathcal{N}}\overline{\mathcal{N}}\overline{\mathcal{N}}$ is
 allowed only if $m=\frac{N}{3} \cdot \mathbb{Z}$. This together with
 $p=\frac{N}{3} \cdot \mathbb{Z}$ shows that the cubic term arises only for
 $\mathds{Z}_3$-symmetries.
\item  \underline{$\overline{\mathcal{N}}\overline{\mathcal{N}}
\overline{\mathcal{N}}\overline{\mathcal{N}}$:} Here we obtain the discrete
charge $4(p-m)$. To find the Dirac-DGSs that allow this quartic term, we
multiply the corresponding condition $4(p-m){=}N \cdot \mathbb{Z}$
by three and apply Eq.~(\ref{pconstraint}): 
\begin{eqnarray*}
\underbrace{4\cdot 3p}_{=4N\cdot\mathbb{Z}} -\; 12m ~=~ 3N \cdot \mathbb{Z} \ ,~~~~\Longrightarrow~~~~ 
12m ~=~ N \cdot \mathbb{Z} \ .
\end{eqnarray*}
Depending on whether $N$ is divisible by 2, 3 and/or 4, we get the conditions
\begin{eqnarray*}
p&\in&\left\{0\,,\, \frac{N\cdot \mathbb{Z}}{3} \right\} \, ,\\
m&\in&\left\{0\,,\, \frac{N\cdot \mathbb{Z}}{2}\,,\, \frac{N\cdot
    \mathbb{Z}}{3} \,,\, \frac{N\cdot \mathbb{Z}}{4} \,,\, \frac{N\cdot
    \mathbb{Z}}{6} \,,\, \frac{N\cdot \mathbb{Z}}{12} \right\} \, .\\ 
\end{eqnarray*}
After rescaling, the only $\mathds{Z}_N$-symmetries that have the potential to
allow the term $\overline{\mathcal{N}}\overline{\mathcal{N}} 
\overline{\mathcal{N}}\overline{\mathcal{N}}$ are those with $N=2,3,4,6,12$. 
However, excluding $M_p$, $B_3$ and $P_6$, one can show explicitly that, of
the remaining $0+3+1+3+8$ possible Dirac-DGSs,  
only three symmetries allow this quartic term: $(4;1,0,0)$, $(12;1,0,4)$ and
$(12;5,0,8)$, see also Appendix~\ref{gutapp}. 
\end{itemize}
Table~\ref{tablepsets} summarizes all lepton and/or baryon number violating 
operators that can possibly occur in Dirac-DGSs. We list all operators
contained in the sets explicitly using
Eqs.~(\ref{BVLVoperators},\ref{BVLVwithN}). The discrete charges of the  
operators are given as well as the DGSs which allow their presence in the
theory. Interestingly, $QQQL$ and its ``friends'' are allowed for
every $\mathds{Z}_N$-symmetry with $N$ {\it not} being a multiple of three;
with respect to proton-decay, these Dirac-DGSs therefore experience the same
shortcoming as $M_p$. In order to get rid of the $QQQL$ and thus stabilize the
proton, we must demand symmetries where $N=3\cdot \mathbb{Z}$ and
$p=\frac{N}{3},\frac{2N}{3}$, the first being $(3;0,0,1)$, $(3;1,0,2)$,
$(6;1,0,2)$, $(6;3,0,2)$. 
A list of all $\mathds{Z}_N$-symmetries up to
$N=14$, showing explicitly the allowed lepton and/or baryon number violating
operators, can be found in Appendix~\ref{gutapp}. For the sake of
completeness, we also include $M_p$, $B_3$ and $P_6$ in this list.
\begin{table}[t]
\begin{center}
\hspace{0mm}\begin{tabular}{||c||c|c|c|c||} \hline \hline 
\begin{tabular}{c}  operators \\ within a set \end{tabular} $\phantom{\Big|}$ &  
$\begin{array}{c}
\overline{U}\overline{D}\overline{D}^{\phantom{|}}  \\ 
QQQH_d\phantom{\Big|}\\
QQ\overline{D}^\dagger\phantom{\Big|} 
\end{array}$ & 
$\begin{array}{c}
QQQL\phantom{\Big|}  \\ 
\overline{U}\overline{U}\overline{D}\overline{E}\phantom{\Big|} \\
\overline{U}\overline{D}\overline{D}\overline{\mathcal{N}}\phantom{\Big|} 
\end{array}$ & 
$\overline{\mathcal{N}}\overline{\mathcal{N}}\overline{\mathcal{N}}$ &
$\overline{\mathcal{N}}\overline{\mathcal{N}}\overline{\mathcal{N}}\overline{\mathcal{N}}$
\\ \hline\hline
\begin{tabular}{c} \\discrete \\ charge \\~\end{tabular} & $\pm m$ &
$\mp p$ & $3(p-m)$ & $4(p-m)$ \\ \hline 
  \begin{tabular}{c} Dirac-DGSs \\ $(N;m,0,p)$ \\ which allow \\ these terms
  \end{tabular} &
$(3;0,0,1)$ & 
  \begin{tabular}{c} 
  all $\mathds{Z}_N\phantom{\Big|}$ \\ with $N\neq 3 \cdot \mathbb{Z}\phantom{\Big|}$ \\
  \hline  those $\mathds{Z}_N\phantom{\Big|}$ \\ with $N=3 \cdot \mathbb{Z}$ \\ and
  $p=0\phantom{\Big|}$ 
  \end{tabular} &
$\begin{array}{c} (3;0,0,1)\phantom{\Big|} \\ (3;1,0,0)\phantom{\Big|} \\
  (3;1,0,2)\phantom{\Big|} \end{array} $ & 
$\begin{array}{c} (4;1,0,0)\phantom{\Big|} \\ (12;1,0,4)\phantom{\Big|} \\
  (12;5,0,8)\phantom{\Big|} \end{array} $ \\\hline \hline 
\end{tabular}
\end{center}
\caption{\label{tablepsets}The lepton and/or baryon number violating operators
 occurring with Dirac-DGSs. The discrete charges of the operators are given as
 well as the symmetries that allow these terms.}
\end{table}

%%%%%%%%%%%%%%%%%%%%%%%%%%%%%%%%

%%%%%%%%%%%%%%%%%%%%%%%%%%%%%%%%

%%%%%%%%%%%%%%%%%%%%%%%%%%%%%%%%

\section{\label{GUTcomp}GUT compatibility}
\cleqn

In this section we analyze the compatibility of the $\mathds{Z}_N$-symmetries
with various grand unified theories (GUTs).\footnote{A similar analysis in
which the $\mathds{Z}_{N\leq12}$-symmetry is required to have $M_p$ as a
subgroup can be found in Ref.~\cite{Mohapatra:2007vd}.} 
Our starting assumption is that the gauge structure of the theory includes $
U(1)_X ~ \times ~G_{\mathrm{GUT}}$
where $G_{\mathrm{GUT}}$ is the gauge group of the chosen GUT, and the $U(1)_X$
factor generates the low-energy discrete symmetry.\footnote{We point out that 
this is a
simplifying assumption, since additional $U(1)^\prime$ factors can arise when
$G_{\mathrm{GUT}}$ breaks down to the Standard Model gauge group
$G_{\mathrm{SM}}$, {\it e.g.} $SO(10) \rightarrow SU(5) \times U(1)^\prime
\rightarrow SU(3)_C \times SU(2)_W \times U(1)_Y \times U(1)^\prime$. In such a
scenario, a {\it  combination} of this $U(1)^\prime$ with the $U(1)_X$  could 
be responsible for the emergence of the discrete symmetry. Here we
shall however assume that the origin of the DGS is independent
of the GUT gauge group.} We therefore get 
$\mathds{Z}_N ~ \times ~G_{\mathrm{GUT}}$.
This structure constrains the possible $\mathds{Z}_N$-symmetries because it
requires all the fields of one $G_{\mathrm{GUT}}$ multiplet to have the same
discrete charge. Note however that it is well possible to have a GUT-compatible
DGS arising from a GUT-incompatible $U(1)_X$; for an example see
Section~\ref{toym}. From the low-energy point of view, the discrete charges are
not uniquely fixed for a specific DGS given in terms of $(N;m,n,p)$. This
ambiguity is parameterized by the integer $r=0,...,N-1$ in Eq.~(\ref{withr})
and can be exploited to find GUT compatible DGSs. 

In the following, we discuss the constraints on the discrete charges for
various GUT(-like) scenarios and their implication for the (non-)existence of the
lepton and 
baryon number violating operators in the set $QQQL$, see Table~\ref{tablepsets}.
\begin{itemize}
\item \underline{$SO(10)$:} The $\bf{16}$ of $SO(10)$ contains all quarks and
  leptons \cite{Fritzsch:1974nn}. Therefore this GUT group requires
$
z_Q ~=~z_{\overline{D}}
~=~z_{\overline{U}}~=~z_L~=~z_{\overline{E}}~=~z_{\overline{\mathcal{N}}}$.
Imposing these relations on Eq.~(\ref{withr}) and setting $n=0$, we
arrive at the necessary conditions for a GUT compatible Dirac-DGS
\begin{equation}
p~=~0\ ,~~~~~m+r~=~0~\mathrm{mod}~N\ ,~~~~~4r~=~0~\mathrm{mod}~N \ .
\end{equation}
Since $p=0$, the operators in the set $QQQL$ are always allowed. Note that
this statement does not depend on the constraints of the anomaly conditions
but only on the presence of the $\mu$-term ($n=0$).
After rescaling, there are actually only two $\mathds{Z}_N$-symmetries
$(N;m,n,p;r)$  which are $SO(10)$ compatible: $(2;1,0,0;1)={M_p}^\prime$ (not
a Dirac-DGS), {\it cf.} Table~\ref{commonDS}, as well as $(4;1,0,0;3)$, {\it
  cf.} Sect.~\ref{toym}. 
%Only the latter is a Dirac-DGS, because matter parity~${M_p}^\prime$ allows
%the Majorana mass term $\overline{\mathcal{N}}\overline{\mathcal{N}}$. 
Note that $SO(10)$ compatible $\mathds{Z}_N$-symmetries are, of course, also
compatible with the following GUTs.  
\item \underline{Georgi-Glashow $SU(5)$} 
[with a possible extra $U(1)^\prime$-factor  with charges  $-1,3,-1,3,-1,-5,-2,2$ for the superfields $Q$, $\overline{D}$, $\overline{U}$, 
$L$, $\overline{E}$, $\overline{\mathcal{N}}$, $H_d$, $H_u$]: 
Here, the $\bf{10}$ decomposes into  $Q,\overline{U},\overline{E}$, and the $\overline{\bf{5}}$ into
  $\overline{D},L$; the right-handed neutrino $\overline{\mathcal{N}}$ lives
  in a singlet of $SU(5)$ \cite{Georgi:1974sy}. So $
z_Q =z_{\overline{U}}=z_{\overline{E}},~
z_{\overline{D}}=z_L$,
leads to the conditions
\begin{equation}
p~=~0\ ,~~~~~~~ m+5r~=~0~\mathrm{mod}~N\ .
\end{equation}
Again, the $QQQL$-set is always allowed in this case.
The $SU(5)$ compatibility of all $\mathds{Z}_N$-symmetries up to 
$N=14$ is shown in  Appendix~\ref{gutapp}, stating explicitly the required
values for $r$.
It is easy to see that there exist
infinitely many such DGSs: Consider for instance $(N;N-5,0,0;1)$ and $N$ being
prime; then no rescaling is possible, so that there are at least as many
$SU(5)$ DGSs as there are prime numbers.

\item \underline{Flipped $SU(5)\times U(1)^{\prime\prime}$} 
[the $U(1)^{\prime\prime}$ factor with charges
  $-1, -1, 3, 3, -5, -1$, $2$, $-2$]: The embedding of the particles into the
multiplets of flipped $SU(5)$ is similar to Georgi-Glashow $SU(5)$. One simply  
switches  ``up'' and ``down'' for the $SU(2)_W$ singlets 
($\overline{D}\leftrightarrow\overline{U}$, 
$\overline{E}\leftrightarrow\overline{\mathcal N}$, 
furthermore $H_d\leftrightarrow H_u$). 
Thus we have 
${\bf{10}}  \rightarrow Q,\overline{D},\overline{\mathcal{N}}$ and 
$\overline{\bf{5}}  \rightarrow \overline{U}, L$; the right-handed electron
  $\overline{E}$ is in the singlet representation of  $SU(5)$ and is charged
  only under $U(1)^{\prime\prime}$, see Ref.~\cite{Barr:1981qv} and also 
Refs.~\cite{Derendinger:1983aj,Antoniadis:1987dx}. This yields 
$z_Q=z_{\overline{D}}=z_{\overline{\mathcal{N}}},~z_{\overline{U}}=z_L$  
leading to
\begin{equation}
p~=~0\ ,~~~~~~~ m+r~=~0~\mathrm{mod}~N\ .
\end{equation}
Also in this case, $QQQL$ cannot be forbidden by a DGS. As with $SU(5)$, there
is an infinite number of flipped $SU(5)$ compatible DGSs.
\item \underline{Pati-Salam [$SU(4) \times SU(2)_W \times SU(2)_R$]:} The
    $({\bf 4},{\bf 2},{\bf 1})$ representation contains the fields $Q,L$,
    while the  $({\overline{\bf 4}},{\bf 1},{\bf 2})$ decomposes into
    $\overline{D},\overline{U},\overline{E},\overline{\mathcal{N}}$
    \cite{Pati:1973uk}. [$SU(3)_C$ comes from  $SU(4)$, while $U(1)_Y$ stems
    from $SU(4)\times SU(2)_R\,$.] Hence 
$
z_Q =z_L,~ 
z_{\overline{D}}=z_{\overline{U}}=z_{\overline{E}}= 
z_{\overline{\mathcal{N}}},$
and then
\begin{equation}
p+4r~=~0~\mathrm{mod}~N\ ,~~~~~~~ 2m+6r~=~0~\mathrm{mod}~N\ . \label{pati}
\end{equation}
As $p$ is not automatically zero, the operators in the set $QQQL$ can be
forbidden by Pati-Salam compatible $\mathds{Z}_N$-symmetries. Actually, there
are only four DGSs which {\it allow} $QQQL$, namely $(2;1,0,0;0)={M_p}$,
$(2;1,0,0;1)={M_p}^\prime$ (both not Dirac-DGSs), and $(4;1,0,0;1)$,
$(4;1,0,0;3)$; all other Pati-Salam  $\mathds{Z}_N$-symmetries forbid
the operators of the set $QQQL$. \\
Interestingly, the number of such DGSs is
finite. We have just stated that with $p=0$, {\it i.e.} allowing $QQQL$,
there are only two Dirac-DGSs. Let us therefore consider
$p=\frac{N}{3}$. Multiplying the first condition of Eq.~(\ref{pati}) by three
and the second by two, we get $N + 12r=3aN\,,~ 4m+12r=2bN$,
with unspecified integers $a,b$. Subtracting the first equation from the
second and solving for $m$~yields
\beq
m~=~ \left( 2b -3a +1   \right) \cdot \frac{N}{4} \ .\label{m-pati}
\eeq
In the case of $N$ not being a multiple of 4, the parameters $a$ and $b$ have
to be chosen such that $m$ is an integer. $p=\frac{N}{3}$ and
Eq.~(\ref{m-pati}) give rise to DGSs of the form (we neglect the value for
$r$)  
\begin{eqnarray*}
&& \left(N~;~ \left( 2b -3a +1   \right) \! \cdot\!\frac{N}{4}~,~0~,~
  \frac{N}{3}\right)  \\[2mm]
&\Longleftrightarrow &\left(12\! \cdot\! \frac{N}{12}~;~ 3 \left( 2b -3a +1
  \right)\! \cdot\! \frac{N}{12}~,~0~,~  4 \!\cdot\!\frac{N}{12}\right) \\[2mm]
&\Longleftrightarrow & \Big(12~;~ 3 \left( 2b -3a +1 \right)~,~0~,~  4 \Big)\ .
\end{eqnarray*}
%(For this step we used $(4|N)$; if this is not the case, we can perform
%analogous steps, arriving at $\mathds{Z}_6$- and  $\mathds{Z}_3$-symmetries.) 
In the last step we have rescaled all parameters with the common factor
$\frac{N}{12}$, which, in general, need not be an integer. Further rescaling
might be possible, depending on the values of $a$ and $b$. For
$p=\frac{2n}{3}$ we obtain a similar result. This shows that Pati-Salam
compatible $\mathds{Z}_N$-symmetries are only possible for $N\leq
12$. Explicit counting yields 9 such Dirac-DGSs, of which $9-2=7$ forbid
$QQQL$. 
\end{itemize}
Summarizing the above results, we have one $SO(10)$, an infinite
number of (flipped) $SU(5)$ and nine Pati-Salam compatible Dirac-DGSs.
The GUT compatibility of all $\mathds{Z}_N$-symmetries with $N\leq 14$
is given in Appendix~\ref{gutapp}. Almost all of them allow
the operators of the set $QQQL$; in order to have proton-decay at an
experimentally acceptable rate it is thus necessary to suppress the term
$QQQL$ in these scenarios. 
Only seven Pati-Salam compatible DGSs forbid the set $QQQL$.
In Appendix~\ref{gutapp} we also give the other allowed sets of lepton
and/or baryon number violating operators discussed in Section~\ref{physicssetc}. 

\section{\label{toym}An example}
\cleqn
To illustrate how  a Dirac-DGS arises from a $U(1)_X$ gauge symmetry, and
how distinct theories can be related by a hypercharge shift,
and how these related theories give rise to different
GUT-compatibilities, we consider three different sets of 
$U(1)_X$-charges to begin  with:

\begin{center}
\begin{tabular}{||c||r|r|r|r|r|r|r|r||}
\hline\hline
Model/Charges & $\phantom{\Big|} X_{Q}\phantom{\Big|} $  &  $X_{\overline{D}}$ & $X_{\overline{U}}$ & $X_{L}$ & $X_{\overline{E}}$ & $X_{\overline{\mathcal{N}}}$ & $X_{H_d}$ & $X_{H_u}$ \nonumber\\
\hline\hline
1 & $\phantom{\Big|} 0\phantom{\Big|} $  & $-3$  & $3$  & $0$  & $-3$  & $3$  & $3$ &  $-3$\nonumber\\\hline
2 & $\phantom{\Big|} 3\phantom{\Big|} $  & $3$  & $-9$  & $-9$  & $15$  & $\phantom{\Big|} 3\phantom{\Big|} $  & $-6$  & $6$\nonumber\\\hline
3 & $\phantom{\Big|} 1\phantom{\Big|} $  & $-1$  & $-1$  & $-3$  & $3$  & $3$   & $0$   & $0$\nonumber\\\hline\hline
\end{tabular}
\end{center}

\noindent These three sets are all free of anomalies and mutually 
related   by $Y$-shifts.\footnote{In fact, they can be 
calculated from Eq.~(8.5) of Ref.~\cite{Dreiner:2005rd} with
$C_1=1$ and  $C_2= -1, 2, 0$.} We assume a vectorlike 
pair of $A$-fields: $X_{A_1}= - 4$, $X_{A_2}=4$.
Then, after $U(1)_X$-symmetry breaking, we get a $\mathds{Z}_4$-symmetry
which might be called matter tetrality~$M_4=(4;1,0,0)$:

\begin{center}   
\begin{tabular}{||c||r|r|r|r|r|r|r|r||}
\hline\hline
Model/Charges & $\phantom{\Big|} z_{Q}\phantom{\Big|} $  &  $z_{\overline{D}}$ & $z_{\overline{U}}$ & $z_{L}$ & $z_{\overline{E}}$ & $z_{\overline{\mathcal{N}}}$ & $z_{H_d}$ & $z_{H_u}$ \nonumber\\
\hline\hline
1 & $\phantom{\Big|} 0\phantom{\Big|} $  & $1$  & $ 3$  & $ 0$  & $ 1$  & $ 3$  & $3$ &  $1$\nonumber\\\hline
2 & $\phantom{\Big|} 3\phantom{\Big|} $  & $ 3$  & $3$  & $3$  & $3$  & $3$  & $2$ &  $2$\nonumber\\\hline
3 & $\phantom{\Big|} 1\phantom{\Big|} $  & $ 3$  & $3$  & $1$  & $ 3$  & $3$  & $ 0$ &  $ 0$\nonumber\\\hline\hline
\end{tabular}
\end{center}

\noindent Therefore, the second model $(r=3$) is compatible with $SO(10)$ and the last
one  ($r=1$) is compatible with Pati-Salam, at least on the discrete level
[but not on the $U(1)_X$-level]. See also Appendix~\ref{gutapp}.

\section{\label{dsco}Conclusion}
\cleqn

When supersymmetrizing the SM, the introduction of a DS to avoid exotic
processes is highly desirable. Such a DS is supposedly 
the remnant of a $U(1)$ broken at high energies. Assuming that the
experimentally observed neutrinos  are Majorana-type, only three
$\mathds Z_N$-symmetries for the MSSM sector are possible: $M_p$, $B_3$ and
$P_6$. Allowing, however, for purely Dirac-type neutrinos (experimentally
still possible), an infinite number of non-equivalent 
discrete anomaly-free $\mathds Z_N$-symmetries is conceivable for the
MSSM$+\overline{\mathcal{N}}$. The existence of the $\mu$-term is a
consequence, not an input, unlike for the three above-mentioned DGSs.  

Up to $N=14$, we have listed all possible DGSs in Appendix~\ref{gutapp},
"decodable" with Eq.~(\ref{withr}). Some of them are compatible with a
GUT-scenario in the sense that the discrete charges are consistent with the
direct product $\mathds Z_N \times G_{\mathrm{GUT}}$. Those DGSs going along
with \mbox{$SO(10)$,} $SU(5)$ and flipped $SU(5)$ automatically allow for
the $QQQL$-set superpotential operators.

\newpage  

Analogously to Table~\ref{commonDS} and in addition to the three  Dirac-DGSs in  Sect.~\ref{toym}, we have collected here five especially
interesting Dirac-DGSs out of the many  in  Appendix~\ref{gutapp}: We show the
explicit charge assignments for all $\mathds Z_{N\leq 6}$-symmetries
$(N;m,n,p;r)$  which forbid $QQQL$. The two trialities can be named
unambiguously according to the classification in
Table~\ref{asasassa}.\footnote{Among the four anomaly-free trialities in the
  table of Appendix~\ref{gutapp}, three forbid $QQQL$ and are hence
  particularly interesting: The Majorana-DGS baryon triality
  $B_3=(3;1,0,1)$ as well as the two Dirac-DGSs lepton triality
  $L_3=(3;0,0,1)$ and matter triality $M_3=(3;1,0,2)$. Since all four
  anomaly-free $\mathds Z_6$-symmetries are matter $N$-alities, we refrain
  from naming the three Dirac-DGSs; the remaining Majorana-DGS is already
  called proton hexality.}
(Note that $M_3$ can be called \emph{flipped} $B_3$, since, barring a
multiplication of the discrete charges by $-1$, $M_3$ is obtained from $B_3$
by flipping the fields in exactly the same way as flipped $SU(5)$ is obtained
from $SU(5)$; likewise, the Dirac-DGS $(6;1,0,2;0)$ could be called \emph{flipped}
$P_6$.\footnote{In general, the flipped version of a DGS with
$(N;m,n,p;r)$ is identical to the symmetry defined by $(N;-m+n-6r,n,p;r)$ [or 
equivalently $(N;m-n+6r,-n,-p;-r)$].
Another example apart from the two above-mentioned ones is 
$(9;1,0,3;0)$ with its flipped version of $(9;1,0,6;0)$.})
When Pati-Salam compatible for a specific 
value for $r$, we give the discrete charges for these cases.  Within the
MSSM-sector, all but the first are as powerful as the aggressive $P_6$ in
Table~\ref{commonDS}.  Regarding the three Majorana-DGSs, {\it i.e.} $M_p$,
$B_3$, and $P_6$, GUT compatibility and the absence of $QQQL$ mutually exclude
each other. 

\vspace{1mm}

\begin{center}
$
\begin{array}{||c||c||c|c|c|c|c|c|c|c|c||}
\hline\hline
\phantom{\Big|} \mbox{Dirac-DGS} \phantom{\Big|} & N & z_Q & z_{\overline{D}} &  z_{\overline{U}} &  z_{L} &
z_{\overline{E}}&   z_{\overline{\mathcal{N}}} & z_{H_d} & z_{H_u} &
\mbox{comment}\\\hline\hline 
\!\!\!\begin{tabular}{c}\mbox{lepton triality$\!\phantom{\Big|}{L_3}^{\!\prime\prime}\!\!$} \\ (3;0,0,1;2) \end{tabular}\!\!\!
  & 3 &  2 & 1 & 1 & 2 & 1 & 1 & 0 & 0 & 
  \!\!\!\!\!\begin{tabular}{c}
  \mbox{the only lepton $N$-ality, }\\ 
  \mbox{allows $\overline{UDD}$ and $\overline{\mathcal{NNN}}$,} \\
  \mbox{Pati-Salam compatible due to $r\!=\!2$} 
  \end{tabular}\!\!\!\!\! \\\hline  
\!\!\!\!\!\!\begin{tabular}{c}\mbox{matter triality$\!\phantom{\Big|}M_3$}\\(3;1,0,2;~e.g.~0)\end{tabular}\!\!\!\!\!\!    & 3 &  0 & 1 & 2 & 1 & 0 & 1 & 2 & 1 & 
  \begin{tabular}{c} \mbox{\,allows
  $\overline{\mathcal{NNN}}\phantom{\Big|}$}\\ 
        \mbox{no $\not \!\! B$ up to dim-5}\end{tabular} \\\hline 
(6;1,0,2;~e.g.~0)& 6& 0 & 1 & 5 & 4 & 3 & 1 & 5 & 1 &
  \begin{tabular}{c}
  \mbox{\,no $\not \!\! L$ and $\not \!\! B$ up to dim-5$\phantom{,\Big|}$}
  \end{tabular}\\\hline
(6;3,0,2;1) & 6 & 1 & 5 & 5 & 1 & 5 & 5 & 0 & 0 &
  \!\!\!\!\!\begin{tabular}{c}
  \mbox{\,no $\not \!\! L$ and $\not \!\! B$ up to dim-5$,\phantom{\Big|}$} \\ 
  \mbox{Pati-Salam compatible due to $r\!=\!1$} 
  \end{tabular}\!\!\!\!\! \\\hline 
(6;3,0,2;4) & 6 & 4 & 5 & 5 & 4 & 5 & 5 & 3 & 3 &
  \!\!\!\!\!\begin{tabular}{c}
  \mbox{\,no $\not \!\! L$ and $\not \!\! B$ up to dim-5$,\phantom{\Big|}$} \\ 
  \mbox{Pati-Salam compatible due to $r\!=\!4$} 
  \end{tabular}\!\!\!\!\!
\\\hline\hline 
\end{array}
$
\end{center}

\vspace{1mm}

\section*{Acknowledgments}
We are grateful for helpful discussions with 
Herbi Dreiner, Hitoshi Murayama,
Pierre Ramond, 
Carlos~A.~Savoy, 
Pierre~Sikivie, 
Patrick~Vaudrevange and
Ak\i{}n~Wingerter.
We both thank the SPhT  at CEA-Saclay and the Physikalisches 
Institut der Universit\"at Bonn for hospitality. M.T. greatly appreciates that he was funded by a Feodor Lynen fellowship 
of the Alexander-von-Humboldt Foundation during 
the initial brainstormy stage of this project while he was a member
of the SPhT at CEA-Saclay. The 
work of C.L. is supported by the University of 
Florida through the Institute for Fundamental 
Theory.

\vspace{10mm}

\begin{appendix}

\noindent{\LARGE{\bf{Appendix}}}

\vspace{-1mm}

\section{\label{endlichregen!}Sets of operators}   
\cleqn
Requiring the existence of the superpotential terms in Eq.~(\ref{mssmsuperpot}) one finds 
that the lepton and/or baryon number violating operators are allowed/forbidden in sets. This
can be shown by rearranging products of superfields. In the following, operators in 
parentheses are allowed by definition, see Eq.~(\ref{mssmsuperpot}). Then the remaining terms always come in
pairs, proving that both are simultaneously either present or
absent. Occasionally, we artificially insert the product of a field and its
complex conjugate, which is trivially invariant under a DGS. So \emph{e.g.}
the existence of $QL\overline{D}$ and necessarily $LH_d\overline{E}$ then requires
the existence of $LL\overline{E}$:
\beqn\label{A0}
QL\overline{D}~\cdot~(LH_d\overline{E})&\sim&LL\overline{E}~\cdot~({QH_d\overline{D}})\  .
\eeqn
In the same manner one finds
\beqn
{QL\overline{D}~\cdot~(H_dH_u)}&\sim&LH_u~\cdot~({QH_d\overline{D}})\ ,\nonumber\\[1mm]  
(QH_d\overline{D})~\cdot~(QH_u\overline{U})~\cdot~(LH_d\overline{E})&\sim&Q\overline{U}\overline{E}H_d~\cdot~QL\overline{D}~\cdot(H_dH_u)\
,\nonumber\\[1mm] 
{QL\overline{D}~\cdot~(H_dH_u)^2}&\sim &L H_u H_d
H_u~\cdot~{(QH_d\overline{D})}\ , \nonumber\\[1mm]
(QH_d\overline{D})~\cdot~(QH_u\overline{U})~\cdot~(L^\dagger
L)&\sim&Q\overline{U}L^\dagger~\cdot~QL\overline{D}~\cdot~(H_dH_u)\ ,\nonumber\\[1mm] 
(QH_d\overline{D})~\cdot~(LH_d\overline{E})~\cdot~({H_u}^\dagger H_u)
&\sim&\overline{E}H_d{H_u}^\dagger~\cdot~{QL\overline{D}~\cdot~(H_dH_u)}\ ,\nonumber\\[1mm] 
{(QH_u\overline{U})~\cdot~(LH_d\overline{E}})~\cdot~(\overline{D}^\dagger\overline{D})&\sim
&\overline{U}\overline{E}\overline{D}^\dagger~\cdot~{QL\overline{D}~\cdot~(H_dH_u)}\
, 
\eeqn
justifying the first ``$\Longleftrightarrow$'' in Eq.~(\ref{BVLVoperators}). Likewise
%%%%%%%%%%%%%%%%%%%
\beqn
(QH_d\overline{D})^2~\cdot~(QH_u\overline{U}) & \sim &
QQQH_d~\cdot~{\overline{UDD}~\cdot~(H_dH_u)}\ ,\nonumber\\[1mm]
(QH_d\overline{D})~\cdot~(QH_u\overline{U})~\cdot~(\overline{D}^\dagger\overline{D})&\sim&
Q Q \overline{D}^\dagger~\cdot~{\overline{UDD}~\cdot~(H_dH_u)}\ ;
\eeqn
%%%%%%%%%%%%%%%%%%%
\beqn
{(QH_d\overline{D})~\cdot~(QH_u\overline{U})^2~\cdot~(LH_d\overline{E}})~\sim~\overline{UUDE}~\cdot~{QQQL~\cdot~(H_dH_u)^2}\
.
\eeqn

\noindent Introducing the right-handed neutrino $\overline{\mathcal{N}}$
and demanding the interaction $LH_u\overline{\mathcal{N}}$, one can similarly
prove the groups of operators given in Eq.~(\ref{BVLVwithN}).
\beqn
LH_u\overline{\mathcal{N}}~\cdot~(LH_d\overline{E})&\sim&LL\overline{E}\overline{\mathcal{N}}~\cdot~(H_dH_u)\
, \nonumber
\\[1mm]
LH_u\overline{\mathcal{N}}~\cdot~(QH_d\overline{D})&\sim&QL\overline{D}\overline{\mathcal{N}}~\cdot~(H_dH_u)\
;
\eeqn
\beqn
(QH_d\overline{D})~\cdot~(LH_u\overline{\mathcal{N}})&\sim&\overline{\mathcal{N}}~\cdot~QL\overline{D}~\cdot~(H_dH_u)\
;
\eeqn
\beqn
(LH_u\overline{\mathcal{N}})~\cdot~(LH_u\overline{\mathcal{N}})&\sim&LH_uLH_u~\cdot~\overline{\mathcal{N}}\overline{\mathcal{N}}\
;
\eeqn
\beqn
(QH_d\overline{D})^2\,\cdot\,(QH_u\overline{U})\,\cdot\,(LH_u\overline{\mathcal{N}})&\sim&
\overline{U}\overline{D}\overline{D}\overline{\mathcal{N}}\,\cdot\,QQQL\,\cdot\,(H_dH_u)^2\
.~~~
\eeqn

%%%%%%%%%%%%%%%%%%%%%%%%%%
%%%%%%%%%%%%%%%%%%%%%%%%%%
%%%%%%%%%%%%%%%%%%%%%%%%%%

\section{\label{stringkacke}The first seven steps of the top-down list in Sect.~\ref{section3}}
\cleqn

Our starting point is an $SU(3)_C \times SU(2)_W \times U(1)_{Y^\prime} \times
U(1)_{X^\prime}$-invariant four-dimensional theory in which the dilaton $S$ has
not yet acquired a vacuum expectation value (VEV). 
Among others, there is the $F$-term 
$  \frac{1}{4}~S~\left(\begin{array}{cc}W_{X^\prime} & W_{Y^\prime}\end{array}\right)\cdot \boldsymbol{K}^\prime\cdot\left(\begin{array}{c} 
W_{X^\prime} \\ W_{Y^\prime} \end{array}\right)$, with $\boldsymbol{K}^\prime$ 
being a $2\times2$ matrix, and  the $D$-terms 
$\overline{\varPhi}~e^{2(V_{X^\prime} {X^\prime}_\varPhi+V_{Y^\prime}
  {Y^\prime}_\varPhi)}~\varPhi$, and 
$
-\frac{1}{2}\ln(S+\overline{S}-{X^\prime}_S V_{X^\prime}-{Y^\prime}_S V_{Y^\prime})
$.
\begin{enumerate}
\item $\boldsymbol{K}^\prime$ has to be positive-definite, and it may be taken 
symmetric. Thus
\beq
\frac{1}{4}~S~\left(\begin{array}{cc}W_{X^\prime} & W_{Y^\prime}\end{array}\right)\cdot\left(\begin{array}{cc} 
{k^\prime}_{11} & {k^\prime}_{12} \\
{k^\prime}_{12} & {k^\prime}_{22} 
  \end{array}\right)\cdot\left(\begin{array}{c} 
W_{X^\prime} \\
W_{Y^\prime} \end{array}\right)\, .\label{dilyuk}
\eeq
\item \label{I} Next we perform an orthogonal transformation to diagonalize
  $\boldsymbol{K}^\prime$. This mixes $W_{X^\prime}$ and  $W_{Y^\prime}$ (and
  equivalently  $V_{X^\prime}$ and  $V_{Y^\prime}$) as well as, for a
  given field~$\varPhi$, its charges ${X^\prime}_\varPhi$ and ${Y^\prime}_\varPhi$. 
\beq
\boldsymbol{K}^\prime ~\rightarrow~ \boldsymbol{K}= \left(\begin{array}{cc} 
k_{X} & 0 \\0 & k_{Y} 
  \end{array}\right)\, .\label{igg}
\eeq 
Thus there is no kinetic mixing anymore between $U(1)_{X}$ and $U(1)_{Y}$. 
This diagonalization is spoiled by the renormalization group evolution;
however, the resulting effects are small. 
$k_{X},k_{Y}$ are called the pseudo Ka\v{c}-Moody levels of $U(1)_X$ 
and $U(1)_Y$.\\
At this point, one might ask the question: {What are the conditions on the original $X^\prime$- and $Y^\prime$-charges such that after a rotation (like in Items~\ref{I},\ref{vii}) the   $X$-charges may be generation-\emph{dependent} whereas the  $Y$-charges are generation-\emph{\underline{in}dependent}?  We have
$$
\left(\begin{array}{c}V_{X^\prime}\\V_{Y^\prime} \end{array}\right)
~\longrightarrow~ 
\left(\begin{array}{c}{V_X}\\{V_Y} \end{array}\right)=
\left(\begin{array}{rr} \cos\gamma & -\sin\gamma\\ \sin\gamma & \cos\gamma  \end{array}\right)\cdot
\left(\begin{array}{c}V_{X^\prime}\\V_{Y^\prime} \end{array}\right)\, ;
$$
$\gamma$ is of course determined by demanding the  $\boldsymbol{K}^\prime$
matrix to be diagonalized (or, in Item~\ref{vii}, that $Y_S$ is rotated away). This rotation gives
$$
\overline{\varPhi}e^{2({X^\prime}_\varPhi V_{X^\prime}+{Y^\prime}_\varPhi
  V_{Y^\prime})}\varPhi 
 \, \longrightarrow\,
\overline{\varPhi}e^{2\big[ \overbrace{(\cos\gamma \cdot
    {X^\prime}_\varPhi-\sin\gamma  \cdot
    {Y^\prime}_\varPhi)}^{\equiv{X_{\varPhi}}} {V_X}+\overbrace{(\sin\gamma
    \cdot {X^\prime}_\varPhi+\cos\gamma \cdot
    {Y^\prime}_\varPhi)}^{\equiv{Y_{\varPhi}}} {V_Y}\big]}\varPhi \ .
$$
Now demand  the resulting $Y$-charges to be generation-independent. Then, 
\emph{e.g.} ${Y}_{Q^i}={Y}_{Q^j}$, leads~to
$$
\sin\gamma \cdot {X^\prime}_{Q^i}+\cos\gamma \cdot {Y^\prime}_{Q^i}=\sin\gamma
\cdot {X^\prime}_{Q^j}+\cos\gamma \cdot {Y^\prime}_{Q^j} \ ,
$$
so that the original $Y$-charges have to fulfill $~
{Y^\prime}_{Q^j}={Y^\prime}_{Q^i}+({X^\prime}_{Q^i}-{X^\prime}_{Q^j})\cdot\tan\gamma$.
}

\item \label{II} Having diagonalized $\boldsymbol{K}$, we investigate the
  effects of a combined $U(1)_X\times U(1)_Y$ gauge-transformation (performed
  \emph{e.g.} to prevent Goldstone bosons), \emph{i.e.} the effects of 
\beqn
\varPhi&\rightarrow& e^{i(\Lambda_X X_\varPhi+\Lambda_Y Y_\varPhi)}\varPhi\ ,\label{phi-trafo}\\
V_{X,Y}&\rightarrow&
  V_{X,Y}-\frac{i}{2}(\Lambda_{X,Y}-\overline{\Lambda}_{X,Y})\ ,\nonumber\\
S&\rightarrow& S-\frac{i}{2}X_S \Lambda_X - \frac{i}{2} Y_S
  \Lambda_Y\ .\label{s-trafo}
\eeqn
Here, the gauge transformation is parameterized by $\Lambda_{X,Y}$. 
The real-valued quantity denoted as $X_S$ is usually written as
$\delta_{\mathrm{GS}}^X$. 
\begin{enumerate}
\item
Eq.~(\ref{phi-trafo}) causes  anomalies (as for the vanishing kinetic mixing terms, there are no mixed terms like $W_XW_Y$) 
\beqn
\frac{1}{32\pi^2}\Bigg[\lambda_X  
\Big(\mathcal A_{CCX}F_C\widetilde{F}_C+\mathcal
A_{WWX}F_W\widetilde{F}_W+\mathcal A_{YYX}F_Y\widetilde{F}_Y+\mathcal
A_{XXX}F_X\widetilde{F}_X  \Big)   \Bigg]\ ,~~~\label{anooomaly}
\eeqn
and the same with the replacements $\lambda_X\rightarrow \lambda_Y$, $\mathcal
A_{..X}\rightarrow \mathcal A_{..Y}$,
plus the  anomalies with gravitation, with \emph{e.g.} $\mathcal
A_{GGX}=\mathrm{Trace} ~ T_X= \frac{1}{2}\sum_i X_i$.  The $\lambda_{X,Y}$ are
the scalar components of $\Lambda_{X,Y}$, and the $\mathcal A_{abc} =
\mathrm{Trace}[\{ T^a,\,T^b  \} \cdot T^c]$ are the anomaly coefficients. The
gauge group generators $T^a$ are assumed to be according to the standard
GUT-convention, so that {\it e.g.}
\beq\label{c-anom}
\mathcal A_{CCX}=\frac{1}{2}\sum_{i=1}^{N_f} (2X_{Q^i}+X_{\overline{U^i}}+
X_{\overline{D^i}}  )+\mathcal A_{CCX}^{\mathrm{beyond~MSSM}}\ ,
\eeq
where $N_f$ is the number of families.
\item 
Eq.~(\ref{s-trafo}) together with Eqs.~(\ref{dilyuk},\ref{igg}) gives
$$
-\frac{i}{8}(X_S\Lambda_X+Y_S\Lambda_Y)~\left(\begin{array}{cc}W_X & W_Y\end{array}\right)\cdot\left(\begin{array}{cc} 
k_{X} & 0 \\
0 & k_{Y} 
  \end{array}\right)\cdot\left(\begin{array}{c} 
W_{X} \\
W_{Y} \end{array}\right)\, ,
$$
which produces
\beqn
-\frac{1}{16}\Bigg[ \lambda_X \, X_S 
\Big(k_CF_C\widetilde{F}_C+k_W F_W\widetilde{F}_W+k_Y F_Y\widetilde{F}_Y+k_X
F_X\widetilde{F}_X  \Big)   \Bigg] \ ,~~~
\eeqn
and the same with the replacements $\lambda_X \rightarrow \lambda_Y$, $X_S
\rightarrow Y_S$
plus the  shifts with gravitation. 
\end{enumerate}

\item  The anomalies are required 
to be canceled by the dilaton-shifts, \emph{i.e.} Items \ref{II}a) and
\ref{II}b) mutually eliminate each other; this is the four-dimensional 
version of the 
Green-Schwarz mechanism \cite{Green:1984sg}, see also \cite{Blumenhagen:2005ga}. Thus it is ensured that the theory is gauge-invariant, \emph{i.e.} one demands
the following anomaly conditions, see {\it e.g.}
Refs.~\cite{Ramond:1995mb,Maekawa:2001uk,Babu:2003zz}, (and the same for $X
\leftrightarrow Y$)  
\beqn
2\pi^2X_S&=&\frac{\mathcal A_{XXX}}{k_X}=\frac{\mathcal
  A_{YYX}}{k_Y}=\frac{\mathcal A_{CCX}}{k_C}=\frac{\mathcal
  A_{WWX}}{k_W}=\frac{\mathcal A_{GGX}}{12}\
.\label{x-eq}
\eeqn
  
\item \label{IV} We let the dilaton acquire a VEV, $
S\rightarrow S+\langle S \rangle$. So:
\begin{itemize}
\item $-\frac{1}{2}\ln(S+\overline{S}-X_SV_X-Y_SV_Y)$ gives, with $\frac{S+\overline{S}-X_SV_X-Y_SV_Y}{2\Re[ \langle S\rangle ]}$ being small, $-\frac{1}{2}\ln(2\Re[\langle S\rangle])-({S+\overline{S}-X_SV_X-Y_SV_Y})/({4\Re[\langle S\rangle]})$,
producing an effective $D$-term $2\xi_XV_X+2\xi_YV_Y$ with
\beq
\xi_X=\frac{X_S}{8\Re[\langle S \rangle]}\ ,~~~~\xi_Y=\frac{Y_S}{8\Re[\langle S
  \rangle]}\ .\label{fieff}
\eeq
This is the Dine-Seiberg-Wen-Witten-mechanism \cite{Dine:1986zy,Dine:1987bq,Atick:1987gy,Dine:1987gj}. 

\item From, \emph{e.g.} $\frac{1}{4}k_CSW_CW_C$, we obtain the gauge
  kinetic terms and thus the gauge coupling constants. Using standard
  GUT-conventions and identifying 
  $2k_C\Re[\langle S \rangle]{=}2/{g_C}^2$, we find, with
  $g_{string}\equiv1/\sqrt{2\Re[\langle S \rangle]}$,
\beq\label{g^2_mal_k}
g_C^2k_C=g_W^2k_W=g_Y^2k_Y=g_X^2k_X=2{g_{string}}^2\ .
\eeq
\end{itemize}     
{}From Eqs.~(\ref{x-eq},\ref{fieff}) and the relation above one finds that
\emph{e.g.} $\xi_X=\frac{{g_{string}}^2~\sum_i X_i}{192\pi^2}$. 

\item \label{vi} Now that $S$ has undergone the gauge shift (Item~\ref{II})
and having acquired a VEV (Item~\ref{IV}), we soak up the constant coefficient
of the $W_{...}W_{...}$, so that, \emph{e.g.}, $W_CW_C$ produces the kinetic
term  $\frac{1}{4}F_CF_C$ rather than $\frac{1}{4{g_C}^2}F_CF_C$. Item~\ref{I}
and Item~\ref{vi} together are called the 'canonicalization of the kinetic
terms of $V_C$'.  

\item \label{vii} Next, we perform yet another orthogonal transformation which
  leaves the freshly canonicalized kinetic terms invariant. This
  transformation rotates away $Y_S$, thus rendering $U(1)_Y$ non-anomalous; 
  so now we have 
$
\mathcal A_{XXY}=0$, also written as $\mathcal A_{YXX}=0$,  and 
$\mathcal A_{CCY}=\mathcal A_{WWY}=\mathcal A_{YYY}=\mathcal A_{GGY}=0$.
\end{enumerate}

%%%%%%%%%%%%%%%%%%%%%%%%%%
%%%%%%%%%%%%%%%%%%%%%%%%%%
%%%%%%%%%%%%%%%%%%%%%%%%%%

\section{\label{ggT}The $\boldsymbol{X}$-charge of the ``effective
$\boldsymbol{\mathfrak{A}}$''}
\cleqn

In the following, we discuss the scenario with two $A$-type particles  
$A_i$ ($i=1,2$). For simplicity, we assume that their charges $X_{A_i}$  
 are positive integers; the generalization to negative
$X$-charges is straightforward. After the breakdown of $U(1)_X$ the effective
operators in the Lagrangian can only have an overall $X$-charge of the form
\beq
X_{\mathrm{total}} ~=~ -~ a_1 \cdot X_{A_1} ~-~ a_2 \cdot X_{A_2}\ ,\label{total}
\eeq
with $a_i \in \mathbb{N}$ for superpotential terms and $a_i \in
\mathbb{Z}$  for K\"ahler potential terms. Notice that, in principle,
operators in the K\"ahler potential can be converted to effective operators
in the superpotential via the Giudice-Masiero/Kim-Nilles mechanism
\cite{Giudice:1988yz,Kim:1994eu}. If the two $X_{A_i}$ have 
a greatest common divisor $d$, we can define new integers $x_{A_i} \equiv
X_{A_i} / d$. With this,  Eq.~(\ref{total}) can be rewritten as
\beq
X_{\mathrm{total}} ~=~ -~ d \cdot \left[\, a_1 \cdot x_{A_1}  ~+~ a_2 \cdot
  x_{A_2} \right] \, .\label{totF}
\eeq
Evidently, $X_{\mathrm{total}}$ is a multiple of $d$. If the square bracket
is not restricted to any subset of $\mathbb{Z}$, we will end up 
with a $\mathds Z_d$-symmetry after $U(1)_X$-breaking. 

The question however remains whether the square bracket can actually take {\it
  any} integer value. To answer this, we first decompose $x_{A_i}$ into prime
  factors~${\xi^{(i)}}_{\alpha}$:  
$$
x_{A_i} = \prod_{\alpha} {\xi^{(i)}}_{\alpha} \ . 
$$
Since $x_{A_1}$ and $x_{A_2}$ do not have a common divisor, one necessarily
has that ${\xi^{(1)}}_\alpha ~ \neq ~ {\xi^{(2)}}_\beta  ,
~\mathrm{for~all}~\alpha,\beta$. Thus the least common multiple of both
$x_{A_i}$ is just their product $x_{A_1}\cdot x_{A_2}$. If one can obtain any
integer within the interval  $[0,x_{A_1} \!\cdot\!x_{A_2}[$  with an
appropriate integer-valued linear combination of the $x_{A_i}$, then the
square bracket in Eq.~(\ref{totF}) can  take any integer value whatsoever.
To check this, we  consider the two linear combinations 
$$
0 ~\leq  a_1 \cdot x_{A_1} ~+~ a_2 \cdot x_{A_2}  < ~ x_{A_1}\cdot x_{A_2}\ ,
$$ 
$$
0 ~\leq  \!\; b_1 \cdot x_{A_1} ~+~\!\; b_2 \cdot x_{A_2} < ~ x_{A_1}\cdot
x_{A_2}\ ,
$$
with $a_2, b_2 \in \{0,1,...,  x_{A_1}-1 \}$ and $a_1,b_1 \in \mathbb{Z}$ such
that the linear combinations of $x_{A_1}$ and $x_{A_2}$ lie within the given
interval. Assuming $a_2  \neq b_2$, we can show that the two linear
combinations can never be matched within the interval $[0,x_{A_1}
\!\cdot\!x_{A_2}[$, since $a_1 \cdot x_{A_1} + a_2 \cdot x_{A_2} = b_1 \cdot
x_{A_1} + b_2 \cdot x_{A_2}$ can be rewritten as 
$$
(a_2~-~ b_2) \cdot x_{A_2} ~=~ (b_1  ~-~ a_1 ) \cdot x_{A_1} \ .
$$
The factor $(a_2-b_2)$ must therefore be a multiple of $x_{A_1}$, which 
however is not the case for $a_2 \neq b_2$ and $a_2, b_2 \in \{0,1,...,  
x_{A_1}-1 \}$. Hence, two linear combinations of the form $ 0 ~\leq ~ a_1 
\cdot x_{A_1} ~+~ a_2 \cdot x_{A_2} ~< ~ x_{A_1}\cdot x_{A_2}$ always yield
different values for  different~$a_2$. Now there are $x_{A_1}$ different
$a_2$. For each $a_2$ one finds $x_{A_2}$ different possible values for $a_1$
such that the linear combination  lies within the interval $[0,x_{A_1}
\!\cdot\!x_{A_2}[$. Thus we can obtain $x_{A_1}\cdot x_{A_2}$  
{\it different} values within the interval 
$[0,x_{A_1}\!\cdot\!x_{A_2}[$ by integer-valued linear 
combinations of $x_{A_i}$. This finally shows that the square bracket 
in Eq.~(\ref{totF}) can take any integer value. 

Likewise, this argumentation can be applied to cases with any number of
$U(1)_X$-breaking fields $A_i$. The remnant discrete symmetry is a $\mathds
Z_{|X_\mathfrak{A}|}$ with $|X_\mathfrak{A}| \equiv d$, 
the greatest common divisor of all $X_{A_i}$.

\section{\label{singclas}Classification of SM-singlets}
\cleqn
 
In Refs.~\cite{Ibanez:1991hv,Ibanez:1991pr,Dreiner:2005rd} it
was assumed that \emph{all} 
non-MSSM particles, including the singlets~$\Omega$ (see Item~\ref{AundO}),  
are heavy, {\it i.e.} two fields must pair up to allow  
a $\mathds{Z}_N$-invariant mass term after $U(1)_X$-breaking. From this, one
could find some simplifications of the anomaly conditions.
If a massive $\Omega$ has a trilinear coupling with~$LH_u$, \emph{i.e.} the
operator $LH_u\Omega$ is allowed, it is  called a Majorana neutrino  
$\overline{\mathcal{N}}_{\mathrm{Maj}}$. Of course this does not exclude  
\emph{other} $\Omega$s with  discrete charges for which  $LH_u\Omega$ 
is  forbidden -- these $\Omega$s then do not carry lepton number and 
are hence not to be called ``neutrinos''. They can have $X$-charges  which are
half-odd-integer or  integer multiples of $N$; other charges are not possible
since they have to add up to an integer multiple of $N$ in order to be
heavy.\footnote{For simplicity we shall exclude cases like 
two $X$-charges being $3/7\cdot N$   and $4/7\cdot N$. We assume that 
all particles 
within one ``$\Omega$-category'' have to have the same discrete charge.} 
Depending on the $X$-charge of the forbidden term $LH_u\Omega$,
there are three mutually exclusive types of  non-neutrino $\Omega$s:
Case~1 has a DGS such that $LH_u$ and   $LH_uLH_u$ are both allowed,
Case~2 has a DGS such that $LH_u$ is not but    $LH_uLH_u$ is allowed,
Case~3 has a DGS such that $LH_u$ and $LH_uLH_u$ are both not allowed;
see the first two lines of Columns~2 and~3  of Table~\ref{om(eg)a}. 
So in Refs.~\cite{Ibanez:1991hv,Ibanez:1991pr,Dreiner:2005rd} the following
cases were treated: 
\emph{a)} no heavy singlets, 
\emph{b)} $\overline{\mathcal{N}}_{\mathrm{Maj}}$,  
\emph{c)} $\Phi$,  
\emph{d)}~$\overline{\mathcal{N}}_{\mathrm{Maj}}+\Phi$, 
\emph{e)} $\overline{\mathcal{N}}_{\mathrm{Maj}^\prime}$,  
\emph{f)} $\Phi^\prime$, 
\emph{g)} $\overline{\mathcal{N}}_{\mathrm{Maj}^\prime}+\Phi^\prime$,   
\emph{h)} $\Xi$, see also Table~\ref{se7en}.  

The situation becomes even more complex once we admit massless $\Omega$s
(as we necessarily have to do in order to deal with Dirac rather than Majorana
neutrinos), see Table~\ref{om(eg)a}. 
There could in principle be exotic particles which are massless
and do not get a mass at least after $U(1)_X$-breaking. 
%(which is the case for  the Dirac neutrinos): 
One would have no or only little systematics in solving 
the discrete gravitation-anomaly condition if  $\Psi$ 
and/or $\Gamma$ and/or  $\Theta$ and/or $\Theta^\prime$ existed (see Lines~2,5
and~8  in Table~\ref{se7en}) -- 
of course there are solutions to the equations, but they are 
quite arbitrary, depending on which $X$-charges one has chosen. Similar to Item~\ref{cwy}~(f), 
the existence of massless SM-neutral particle spoils the predictability of $\mathcal{A}_{GGX}$.
For that reason we shall not admit these particles in our treatment 
here. [In Ref.~\cite{Jack:2003pb}, the  discrete 
gravitation-anomaly condition is not solved and the singlet
particle content is not specify, so that they effectively work with 
a theory with $\Theta^\prime$ and  $\Phi^\prime$. See also
Appendix~\ref{casestudy}.] 
On the other hand, the analysis of a theory containing Dirac neutrinos
as well as heavy singlets does not differ from the analysis of 
Dirac neutrinos alone, so its results can be taken over wholesale.

\begin{table}[thb]
\vspace{-0mm}
\begin{center}
\begin{tabular}{||c||c|c|c||}
\hline\hline
$\phantom{\bigg|}$ & $\frac{X_{LH_u\Omega}}{N}=\mbox{int.}   $& $\frac{X_{LH_u\Omega}}{N}=\mbox{int.}+\frac{1}{2}$ & $\frac{X_{LH_u\Omega}}{N}\neq\mbox{int.}, \mbox{int.}+\frac{1}{2}$   \\
\hline   \hline
\begin{tabular}{c}\\$\frac{X_\Omega}{N}=\mbox{int.}$\\
$\Rightarrow \exists~\Omega\Omega$ \\~
\end{tabular} & Case~1:~ $\Omega\equiv \overline{\mathcal{N}}_{\mathrm{Maj}}  $& Case~2:~ $\Omega\equiv\Phi^\prime  $ &  Case~3:~ $\Omega\equiv\Xi  $  \\
\hline   
\begin{tabular}{c}\\$\frac{X_\Omega}{N}=\mbox{int.}+\frac{1}{2}$\\
$\Rightarrow \exists~\Omega\Omega$\\~
\end{tabular}                     &  Case~2:~ $\Omega\equiv \overline{\mathcal{N}}_{\mathrm{Maj}^\prime}$ &   Case~1:~ $\Omega\equiv\Phi  $&   Case~3:~ $\Omega\equiv\Xi$  \\
\hline
 \begin{tabular}{c}$\frac{X_\Omega}{N}\neq\mbox{int.},\mbox{int.}+\frac{1}{2}$\\
$\Rightarrow \not\!\exists~\Omega\Omega$
\end{tabular}                      &  Case~3:~$\Omega\equiv \overline{\mathcal{N}}_{\mathrm{Dirac}}$  &   Case~3:~$\Omega\equiv\Psi  $& 
\begin{tabular}{ll}
\\Case~1:&$\!\!\Omega\equiv\Theta $,\\
Case~2:&$\!\!\Omega\equiv\Theta^\prime $,\\
Case~3:&$\!\!\Omega\equiv\Gamma $\\~
 \end{tabular}\\
\hline\hline
\end{tabular}
\end{center}\vspace{-4mm}
\caption{\label{om(eg)a}Classification of different $\Omega$s, with
Case~1 ($\exists LH_u$, $\exists LH_u LH_u$),
Case~2 ($\not\!\exists LH_u$, $\exists LH_u LH_u$),
Case~3 ($\not\!\exists LH_u$, $\not\!\exists LH_u LH_u$).}
\vspace{2mm}
\begin{center}
\begin{tabular}{||c|c|c||}\hline\hline
 $\phantom{\Big|}  $ Case $\phantom{\Big|}  $ & SM-singlet content &  reference \\
\hline \hline
  $\phantom{\Big|}  $ 1 $\phantom{\Big|}  $ & $\overline{\mathcal{N}}_{\mathrm{Maj}}$,  $\Phi$,  $\overline{\mathcal{N}}_{\mathrm{Maj}}+\Phi $ & treated in Refs.~\cite{Ibanez:1991hv,Ibanez:1991pr,Dreiner:2005rd}\\
\hline 
  $\phantom{\Big|}  $  & $\Theta $, $\Phi+\Theta $, $\overline{\mathcal{N}}_{\mathrm{Maj}}+\Theta$,
  $\overline{\mathcal{N}}_{\mathrm{Maj}}+\Theta+\Phi $ & \\
\hline\hline 
 $\phantom{\Big|}  $  2 $\phantom{\Big|}  $ & $\overline{\mathcal{N}}_{\mathrm{Maj}^\prime}$, $\Phi^\prime$, $\overline{\mathcal{N}}_{\mathrm{Maj}^\prime}+\Phi^\prime $ &   treated in Refs.~\cite{Ibanez:1991hv,Ibanez:1991pr,Dreiner:2005rd}\\
\hline 
 $\phantom{\Big|}  $ &  $\Theta^\prime $, $\Phi^\prime+\Theta^\prime $ & examples given in
 \cite{dasewigepaper,Jack:2003pb} \\
\hline 
  $\phantom{\Big|}  $ & $\overline{\mathcal{N}}_{\mathrm{Maj}^\prime}+\Theta^\prime $, $\overline{\mathcal{N}}_{\mathrm{Maj}^\prime}+\Theta^\prime+\Phi^\prime $ & \\
\hline \hline
  $\phantom{\Big|}  $ 3$\phantom{\Big|}  $  & $\overline{\mathcal{N}}_{\mathrm{Dirac}}$, $\overline{\mathcal{N}}_{\mathrm{Dirac}}+\Xi$ & treated here\\
\hline 
 $\phantom{\Big|}  $  & $\Xi$ & treated in Refs.~\cite{Ibanez:1991hv,Ibanez:1991pr,Dreiner:2005rd} \\\hline
& \begin{tabular}{c} $\phantom{\Big|}  $ 
$\Psi$, $\Gamma$, $\Xi+\Psi$, $\Xi+\Gamma$, 
$\overline{\mathcal{N}}_{\mathrm{Dirac}}+\Psi$, 
$\overline{\mathcal{N}}_{\mathrm{Dirac}}+\Gamma $,  $\Psi+\Gamma $, \\  $\phantom{\Big|}  $  $\Xi+\overline{\mathcal{N}}_{\mathrm{Dirac}}+\Gamma  $,  
 $\Xi+\overline{\mathcal{N}}_{\mathrm{Dirac}}+\Psi$, $\Xi+\Psi+\Gamma$, \\$\phantom{\Big|}  $ 
$\overline{\mathcal{N}}_{\mathrm{Dirac}}+\Psi+\Gamma$, $\Xi+\overline{\mathcal{N}}_{\mathrm{Dirac}}+\Psi+ \Gamma  $ \end{tabular}& \\
\hline\hline
\end{tabular}
\end{center}\vspace{-4mm}
\caption{\label{se7en}Mutually different theories and which of these are treated here.}
\end{table}
\vspace{-0mm}

\newpage

\section{\label{casestudy}Case study: a model by Jack, Jones and Wild}
\cleqn
In Ref.~\cite{Jack:2003pb} one is given a model with a non-anomalous $U(1)_X$
(only the mixed anomalies are imposed) and  four $A$-superfields. Explicitly
no right-handed neutrinos are assumed, so tacitly the existence of fields like
$\Theta^\prime$ and/or  $\Phi^\prime$ (see Appendix~\ref{singclas}) must be
assumed to cancel $\mathcal{A}_{GGX}$ and $\mathcal{A}_{XXX}$. The model is of
Case~2, \emph{i.e.} $LH_uLH_u$ is allowed but not so  $LH_u$.   
Their considerations lead to
a set of $X$-charges (note that their $X_{\overline{E^2}}\equiv e_2$ should read $3143/300$ and not $3143/100$) with a free parameter $X_{H_u}=h_2$; if we set
$h_2=3\alpha$, then $\alpha$ is the parameter of a $Y$-shift. We are now going
to extract which discrete symmetry is hidden in these $X$-charges. 
First we rescale all charges by a factor of 2700 so that they are all
integers. Now, the $A$s have charges $-2700, -2700, -720, -234$. The greatest
common divisor is 18, hence we have a~$\mathds{Z}_{18}$.  Then we
pick $h_2 = 2309/900$. Examining the resulting charges mod~18 gives 0, 15, 3,
12, 3;  3, 15   for the fields $Q^i$,   
$\overline{D^i}$, 
$\overline{U^i}$,  
$L^i$,  
$\overline{E^i}$; 
$H_d$, $H_u$. Finally we re-rescale by a factor of three, giving the
discrete charges of $P_6$, see Table~\ref{commonDS}.

\newpage

\section{\label{gutapp}Table of all $\boldsymbol{\mathds{Z}_N}$-symmetries
  with $\boldsymbol{N\leq14}$} 
\cleqn
The following table shows all $\mathds{Z}_N$-symmetries $(N;m,n,p)$ up to
$N=14$, which can be converted into the corresponding discrete charges with
the help of Eq.~(\ref{withr}).  In addition to the Dirac-DGSs, {\it i.e.}
those which forbid the Majorana mass term
$\overline{\mathcal{N}}\overline{\mathcal{N}}$, we list also the standard DGSs
$M_p$, $B_3$ and $P_6$ for completeness. The latter three symmetries allow
$\overline{\mathcal{N}}\overline{\mathcal{N}}$ and thus $LH_uLH_u$, resulting
in Majorana-type light neutrinos. Except for the only baryon $N$-ality $B_3$
(which is not a Dirac-DGS), all symmetries forbid the operators of the set
$QL\overline{D}$. Furthermore, except for the only lepton 
$N$-ality $L_3=(3;0,0,1)$, all Dirac-DGSs forbid $\overline{UDD}$ and are
matter $N$-alities.   
Not all DGSs are compatible with a GUT scenario. However, if they are,
the parameter $r$ in Eq.~(\ref{withr}) has to take specific values which are
given in the table. Furthermore, the sets of lepton and/or baryon number
violating operators allowed by the DGSs are marked with the 
symbol~$\checkmark$.  
\begin{table}[h]
{ %\scriptsize
\vspace{-0mm}
\tiny
\begin{center}
$
\begin{array}{||c||c|c|c|l||} \hline\hline
  (N;m,n,p)   
 & QQQL &  
\phantom{\Big|}\overline{\mathcal{N}}\overline{\mathcal{N}}\overline{\mathcal{N}} &
 \phantom{\Big|}\overline{\mathcal{N}}\overline{\mathcal{N}}\overline{\mathcal{N}}\overline{\mathcal{N}}
 & \text{~~~~~~GUT-compatibility} 
 \\\hline\hline 
 {M_p\!\!\:\!\!\: = (2;1,0,0)}    & \checkmark &  & \checkmark &
      \begin{array}{ll} r=0: & ~~~\mbox{Pati-Salam} \\ r=1: & ~~~SO(10)  \end{array} \\\hline

{B_3 = (3;1,0,1)}    &  & \checkmark & \checkmark &  \\\hline

{P_6 \:\!= (6;5,0,2)} &    &  & \checkmark &  \\\hline \hline

{L_3=(3;0,0,1)}  &  & \checkmark &  & 
      \begin{array}{ll} r=2: & ~~~\mbox{Pati-Salam} \end{array} \\\hline

{(3;1,0,0)}    & \checkmark & \checkmark &  & 
      \begin{array}{ll} r=1: & ~~~SU(5) \\ r=2: & ~~~\mbox{flipped}~SU(5)  \end{array} \\\hline

{M_3\!\!\:\!\!\:=(3;1,0,2)}    &  & \checkmark &  &  \\\hline

{M_4\!\!\:\!\!\:=(4;1,0,0)}    & \checkmark &  & \checkmark & 
      \begin{array}{ll} r=1: & ~~~\mbox{Pati-Salam} \\ r=3: & ~~~SO(10)\end{array} \\\hline

{(5;1,0,0)}    & \checkmark &  &  &
     \begin{array}{ll} r=4: & ~~~\mbox{flipped}~SU(5)  \end{array}  \\\hline

{(5;2,0,0)}    & \checkmark &  &  & 
     \begin{array}{ll} r=3: & ~~~\mbox{flipped}~SU(5)  \end{array} \\\hline

{(6;1,0,0)}    & \checkmark &  &  & 
      \begin{array}{ll} r=1: & ~~~SU(5) \\ r=5: &~~~ \mbox{flipped}~SU(5)  \end{array} \\\hline

{(6;1,0,2)}    &  &  &  &  \\\hline

{(6;3,0,2)}    &  &  &  & 
      \begin{array}{ll} r=1,4: & \mbox{Pati-Salam}   \end{array} \\\hline

{(7;1,0,0)}   & \checkmark &  &  & 
      \begin{array}{ll} r=4: & ~~~SU(5) \\ r=6: & ~~~\mbox{flipped}~SU(5)  \end{array} \\\hline

{(7;2,0,0)}    & \checkmark &  &  & 
      \begin{array}{ll} r=1: & ~~~SU(5) \\ r=5: &~~~ \mbox{flipped}~SU(5)  \end{array} \\\hline

{(7;3,0,0)}    & \checkmark &  &  & 
      \begin{array}{ll} r=4: & ~~~\mbox{flipped}~SU(5) \\   r=5: &~~~ SU(5) \end{array} \\\hline

{(8;1,0,0)}    & \checkmark &  &  & 
      \begin{array}{ll} r=3: & ~~~SU(5) \\ r=7: & ~~~\mbox{flipped}~SU(5)  \end{array}  \\\hline

{(8;3,0,0)}    & \checkmark &  &  & 
      \begin{array}{ll} r=1: & ~~~SU(5) \\ r=5: & ~~~\mbox{flipped}~SU(5)  \end{array} \\\hline

{(9;1,0,0)}    & \checkmark &  &  & 
      \begin{array}{ll} r=7: & ~~~SU(5) \\ r=8: & ~~~\mbox{flipped}~SU(5)  \end{array} \\\hline

{(9;1,0,3)}    &  &  &  &  \\\hline

{(9;1,0,6)}    &  &  &  &  \\\hline

{(9;2,0,0)}    & \checkmark &  &  &  
     \begin{array}{ll} r=5: & ~~~SU(5) \\ r=7: & ~~~\mbox{flipped}~SU(5)  \end{array}\\\hline

{(9;2,0,3)}    &  &  &  &  \\\hline

{(9;2,0,6)}    &  &  &  &  \\\hline

{(9;4,0,0)}    & \checkmark &  &  & 
     \begin{array}{ll} r=1: & ~~~SU(5) \\ r=5: & ~~~\mbox{flipped}~SU(5)  \end{array} \\\hline

{(9;4,0,3)}    &  &  &  &  \\\hline

{(9;4,0,6)}    &  &  &  &  \\\hline

{(10;1,0,0)}    & \checkmark &  &  &  
      \begin{array}{ll}  r=9: & ~~~\mbox{flipped}~SU(5)  \end{array}\\\hline

{(10;3,0,0)}    & \checkmark &  &  &  
      \begin{array}{ll}  r=7: & ~~~\mbox{flipped}~SU(5)  \end{array} \\\hline

{(11;1,0,0)}    & \checkmark &  &  & 
     \begin{array}{ll} r=2: & ~\;\:\!\!SU(5) \\ r=10: &~\;\:\!\! \mbox{flipped}~ SU(5)  \end{array} \\\hline

{(11;2,0,0)}    & \checkmark &  &  & 
      \begin{array}{ll} r=4: & ~~~SU(5) \\ r=9: &~~~ \mbox{flipped}~SU(5)  \end{array} \\\hline

{(11;3,0,0)}    & \checkmark &  &  & 
      \begin{array}{ll} r=6: & ~~~SU(5) \\ r=8: & ~~~\mbox{flipped}~SU(5)  \end{array} \\\hline

{(11;4,0,0)}    & \checkmark &  &  & 
      \begin{array}{ll} r=7: & ~~~\mbox{flipped}~SU(5) \\  r=8: & ~~~SU(5)\end{array} \\\hline

{(11;5,0,0)}    & \checkmark &  &  & 
      \begin{array}{ll} r=6: & ~\,\mbox{flipped}~SU(5) \\  r=10: & ~\,SU(5)\end{array} \\\hline

{(12;1,0,0)}    & \checkmark &  &  & 
      \begin{array}{ll} r=7: & ~\,SU(5) \\ r=11: & ~\,\mbox{flipped}~SU(5)  \end{array} \\\hline

{(12;1,0,4)}    & &  & \checkmark &  \\\hline

{(12;1,0,8)}    &  &  & &  \\\hline

{(12;3,0,4)}    &  &  &  & 
      \begin{array}{ll} r=5,11\!: & \!\!\!\:\mbox{Pati-Salam} \end{array} \\\hline

{(12;3,0,8)}   &  &  &  &  
      \begin{array}{ll} r=1,7: & \mbox{Pati-Salam} \end{array}\\\hline

{(12;5,0,0)}   &  \checkmark & & &
      \begin{array}{ll} r=7: & ~\,\mbox{flipped}~SU(5) \\  r=11: & ~\,SU(5)\end{array} \\\hline

{(12;5,0,4)}    &  &  &  &  \\\hline

{(12;5,0,8)}    &  &  &  \checkmark &  \\\hline

{(13;1,0,0)}    & \checkmark &  &  &  
      \begin{array}{ll} r=5: & ~\,SU(5) \\ r=12: & ~\,\mbox{flipped}~SU(5)  \end{array}\\\hline

{(13;2,0,0)}    & \checkmark &  &  &  
      \begin{array}{ll} r=10: & ~\,SU(5) \\ r=11: & ~\,\mbox{flipped}~SU(5)  \end{array} \\\hline

{(13;3,0,0)}    & \checkmark &  &  &  
      \begin{array}{ll} r=2: & ~\,SU(5) \\ r=10: & ~\,\mbox{flipped}~SU(5)  \end{array} \\\hline

{(13;4,0,0)}   & \checkmark &  &  &  
      \begin{array}{ll} r=7: & ~\:\,\: SU(5) \\ r=9: & ~\:\,\: \mbox{flipped}~SU(5)  \end{array} \\\hline

{(13;5,0,0)}    & \checkmark &  &  & 
      \begin{array}{ll} r=8: & ~\,\mbox{flipped}~SU(5) \\  r=12: & ~\,SU(5)\end{array} \\\hline

{(13;6,0,0)}    & \checkmark &  &  &  
      \begin{array}{ll} r=4: & ~~~SU(5) \\ r=7: & ~~~\mbox{flipped}~SU(5)  \end{array} \\\hline

{(14;1,0,0)}    & \checkmark &  &  &  
      \begin{array}{ll} r=11: & ~\,SU(5) \\ r=13: & ~\,\mbox{flipped}~SU(5)  \end{array} \\\hline

{(14;3,0,0)}    & \checkmark &  &  &  
      \begin{array}{ll} r=5: & ~\,SU(5) \\ r=11: & ~\,\mbox{flipped}~SU(5)  \end{array} \\\hline

{(14;5,0,0)}     & \checkmark &  &  &  
      \begin{array}{ll}   r=9: & ~\,\mbox{flipped}~SU(5) \\ r=13: & ~\,SU(5) \end{array} \\\hline\hline
\end{array}
$
\end{center}
}
\end{table}
\end{appendix}

\bibliographystyle{hunsrt}
\bibliography{references}

\begin{thebibliography}{10}

\bibitem{Strumia:2006db}
A.~Strumia and F.~Vissani.
\newblock 2006, \mbox{hep-ph/0606054}.

\bibitem{Cleaver:1997nj}
G.~Cleaver et~al.
\newblock {\em Phys. Rev.}, D57:2701, 1998, \mbox{hep-ph/9705391}.

\bibitem{Langacker:1998ut}
P.~Langacker.
\newblock {\em Phys. Rev.}, D58:093017, 1998, \mbox{hep-ph/9805281}.

\bibitem{Gogoladze:2001kj}
I.~Gogoladze and A.~Perez-Lorenzana.
\newblock {\em Phys. Rev.}, D65:095011, 2002, \mbox{hep-ph/0112034}.

\bibitem{Hung:2002qp}
P.~Q. Hung.
\newblock {\em Phys. Rev.}, D67:095011, 2003, \mbox{hep-ph/0210131}.

\bibitem{Gherghetta:2003he}
T.~Gherghetta.
\newblock {\em Phys. Rev. Lett.}, 92:161601, 2004, \mbox{hep-ph/0312392}.

\bibitem{Abel:2004tt}
S.~Abel, A.~Dedes, and K.~Tamvakis.
\newblock {\em Phys. Rev.}, D71:033003, 2005, \mbox{hep-ph/0402287}.

\bibitem{Davoudiasl:2005ks}
H.~Davoudiasl, R.~Kitano, G.~D. Kribs, and H.~Murayama.
\newblock {\em Phys. Rev.}, D71:113004, 2005, \mbox{hep-ph/0502176}.

\bibitem{Gabriel:2006ns}
S.~Gabriel and S.~Nandi.
\newblock {\em Phys. Lett.}, B655:141, 2007, \mbox{hep-ph/0610253}.

\bibitem{Nandi:2007cw}
S.~Nandi and Z.~Tavartkiladze.
\newblock 2007, \mbox{arXiv:0708.4033}.

\bibitem{Demir:2007dt}
D.~A. Demir, L.~L. Everett, and P.~Langacker.
\newblock 2007, \mbox{arXiv:0712.1341}.

\bibitem{Dick:1999je}
K.~Dick, M.~Lindner, M.~Ratz, and D.~Wright.
\newblock {\em Phys. Rev. Lett.}, 84:4039, 2000, \mbox{hep-ph/9907562}.

\bibitem{Asaka:2005cn}
T.~Asaka, K.~Ishiwata, and T.~Moroi.
\newblock {\em Phys. Rev.}, D73:051301, 2006, \mbox{hep-ph/0512118}.

\bibitem{Gu:2006dc}
P.-H. Gu and H.-J. He.
\newblock {\em JCAP}, 0612:010, 2006, \mbox{hep-ph/0610275}.

\bibitem{Gu:2007mi}
P.-H. Gu, H.-J. He, and U.~Sarkar.
\newblock {\em JCAP}, 0711:016, 2007, \mbox{arXiv:0705.3736}.

\bibitem{Gu:2007mc}
P.-H. Gu, H.-J. He, and U.~Sarkar.
\newblock {\em Phys. Lett.}, B659:634, 2008, \mbox{arXiv:0709.1019}.

\bibitem{Gu:2007gy}
P.-H. Gu.
\newblock 2007, \mbox{arXiv:0710.1044}.

\bibitem{Lindner:2005as}
M.~Lindner, M.~Ratz, and M.~A. Schmidt.
\newblock {\em JHEP}, 09:081, 2005, \mbox{hep-ph/0506280}.

\bibitem{Hagedorn:2005kz}
C.~Hagedorn and W.~Rodejohann.
\newblock {\em JHEP}, 07:034, 2005, \mbox{hep-ph/0503143}.

\bibitem{Ibanez:1991hv}
L.~E. Ib\'a\~nez and G.~G. Ross.
\newblock {\em Phys. Lett.}, B260:291, 1991.

\bibitem{Ibanez:1991pr}
L.~E. Ib\'a\~nez and G.~G. Ross.
\newblock {\em Nucl. Phys.}, B368:3, 1992.

\bibitem{Dreiner:2005rd}
H.~K. Dreiner, C.~Luhn, and M.~Thormeier.
\newblock {\em Phys. Rev.}, D73:075007, 2006, \mbox{hep-ph/0512163}.

\bibitem{Dreiner:2003hw}
H.~K. Dreiner and M.~Thormeier.
\newblock {\em Phys. Rev.}, D69:053002, 2004, \mbox{hep-ph/0305270}.

\bibitem{Dreiner:2003yr}
H.~K. Dreiner, H.~Murayama, and M.~Thormeier.
\newblock {\em Nucl. Phys.}, B729:278, 2005, \mbox{hep-ph/0312012}.

\bibitem{Harnik:2004yp}
R.~Harnik, D.~T. Larson, H.~Murayama, and M.~Thormeier.
\newblock {\em Nucl. Phys.}, B706:372, 2005, \mbox{hep-ph/0404260}.

\bibitem{Larson:2004ji}
D.~T. Larson.
\newblock 2004, \mbox{hep-ph/0410035}.

\bibitem{Grossman:1998py}
Y.~Grossman and H.~E. Haber.
\newblock {\em Phys. Rev.}, D59:093008, 1999, \mbox{hep-ph/9810536}.

\bibitem{Sakai:1981pk}
N.~Sakai and T.~Yanagida.
\newblock {\em Nucl. Phys.}, B197:533, 1982.

\bibitem{Weinberg:1981wj}
S.~Weinberg.
\newblock {\em Phys. Rev.}, D26:287, 1982.

\bibitem{Allanach:2003eb}
B.~C. Allanach, A.~Dedes, and H.~K. Dreiner.
\newblock {\em Phys. Rev.}, D69:115002, 2004, \mbox{hep-ph/0309196}.

\bibitem{Farrar:1978xj}
G.~R. Farrar and P.~Fayet.
\newblock {\em Phys. Lett.}, B76:575, 1978.

\bibitem{Dimopoulos:1981dw}
S.~Dimopoulos, S.~Raby, and F.~Wilczek.
\newblock {\em Phys. Lett.}, B112:133, 1982.

\bibitem{Bento:1987mu}
M.~C. Bento, L.~J. Hall, and G.~G. Ross.
\newblock {\em Nucl. Phys.}, B292:400, 1987.

\bibitem{Krauss:1988zc}
L.~M. Krauss and F.~Wilczek.
\newblock {\em Phys. Rev. Lett.}, 62:1221, 1989.

\bibitem{Banks:1989ag}
T.~Banks.
\newblock {\em Nucl. Phys.}, B323:90, 1989.

\bibitem{Preskill:1990bm}
J.~Preskill and L.~M. Krauss.
\newblock {\em Nucl. Phys.}, B341:50, 1990.

\bibitem{dasewigepaper}
C.~A. Savoy and M.~Thormeier.
\newblock {\em To be published soon}.

\bibitem{Giudice:1988yz}
G.~F. Giudice and A.~Masiero.
\newblock {\em Phys. Lett.}, B206:480, 1988.

\bibitem{Kim:1994eu}
J.~E. Kim and H.-P. Nilles.
\newblock {\em Mod. Phys. Lett.}, A9:3575, 1994, \mbox{hep-ph/9406296}.

\bibitem{Weinberg:1995mt}
S.~Weinberg.
\newblock {The Quantum Theory of Fields. Vol. 1}.
\newblock Cambridge, UK: Univ. Pr. (1995) 609 p.

\bibitem{Weinberg:1996kw}
S.~Weinberg.
\newblock 1996, \mbox{hep-th/9702027}.

\bibitem{Wichmann:1963}
E.~Wichmann and J.~Crichton.
\newblock {\em Phys. Rev.}, 132:2788, 1963.

\bibitem{Maki:1962mu}
Z.~Maki, M.~Nakagawa, and S.~Sakata.
\newblock {\em Prog. Theor. Phys.}, 28:870, 1962.

\bibitem{Kapetanakis:1992jj}
D.~Kapetanakis, P.~Mayr, and H.-P. Nilles.
\newblock {\em Phys. Lett.}, B282:95, 1992.

\bibitem{Dreiner:2006xw}
H.~K. Dreiner, C.~Luhn, H.~Murayama, and M.~Thormeier.
\newblock {\em Nucl. Phys.}, B774:127, 2007, \mbox{hep-ph/0610026}.

\bibitem{Dreiner:2007vp}
H.~K. Dreiner, C.~Luhn, H.~Murayama, and M.~Thormeier.
\newblock 2007, \mbox{arXiv:0708.0989}.

\bibitem{Mohapatra:2007vd}
R.~N. Mohapatra and M.~Ratz.
\newblock {\em Phys. Rev.}, D76:095003, 2007, \mbox{arXiv:0707.4070}.

\bibitem{Fritzsch:1974nn}
H.~Fritzsch and P.~Minkowski.
\newblock {\em Ann. Phys.}, 93:193, 1975.

\bibitem{Georgi:1974sy}
H.~Georgi and S.~L. Glashow.
\newblock {\em Phys. Rev. Lett.}, 32:438, 1974.

\bibitem{Barr:1981qv}
S.~M. Barr.
\newblock {\em Phys. Lett.}, B112:219, 1982.

\bibitem{Derendinger:1983aj}
J.~P. Derendinger, J.~E. Kim, and D.~V. Nanopoulos.
\newblock {\em Phys. Lett.}, B139:170, 1984.

\bibitem{Antoniadis:1987dx}
I.~Antoniadis, J.~R. Ellis, J.~S. Hagelin, and D.~V. Nanopoulos.
\newblock {\em Phys. Lett.}, B194:231, 1987.

\bibitem{Pati:1973uk}
J.~C. Pati and A.~Salam.
\newblock {\em Phys. Rev.}, D8:1240, 1973.

\bibitem{Green:1984sg}
M.~B. Green and J.~H. Schwarz.
\newblock {\em Phys. Lett.}, B149:117, 1984.

\bibitem{Blumenhagen:2005ga}
R.~Blumenhagen, G.~Honecker, and T.~Weigand.
\newblock {\em JHEP}, 06:020, 2005, \mbox{hep-th/0504232}.

\bibitem{Ramond:1995mb}
P.~Ramond.
\newblock 1995, \mbox{hep-ph/9604251}.

\bibitem{Maekawa:2001uk}
N.~Maekawa.
\newblock {\em Prog. Theor. Phys.}, 106:401, 2001, \mbox{hep-ph/0104200}.

\bibitem{Babu:2003zz}
K.~S. Babu, T.~Enkhbat, and I.~Gogoladze.
\newblock {\em Nucl. Phys.}, B678:233, 2004, \mbox{hep-ph/0308093}.

\bibitem{Dine:1986zy}
M.~Dine, N.~Seiberg, X.~G. Wen, and E.~Witten.
\newblock {\em Nucl. Phys.}, B278:769, 1986.

\bibitem{Dine:1987bq}
M.~Dine, N.~Seiberg, X.~G. Wen, and E.~Witten.
\newblock {\em Nucl. Phys.}, B289:319, 1987.

\bibitem{Atick:1987gy}
J.~J. Atick, L.~J. Dixon, and A.~Sen.
\newblock {\em Nucl. Phys.}, B292:109, 1987.

\bibitem{Dine:1987gj}
M.~Dine, I.~Ichinose, and N.~Seiberg.
\newblock {\em Nucl. Phys.}, B293:253, 1987.

\bibitem{Jack:2003pb}
I.~Jack, D.~R.~T. Jones, and R.~Wild.
\newblock {\em Phys. Lett.}, B580:72, 2004, \mbox{hep-ph/0309165}.

\end{thebibliography}

\end{document}